\documentclass[aps,pre,10pt,twocolumn]{revtex4-2}
\usepackage[colorlinks = True, linkcolor = blue, citecolor = blue, breaklinks=True]{hyperref}
\usepackage{tikz}
\usepackage{pgfplots}
\usepgfplotslibrary{groupplots}
\pgfplotsset{compat=1.18}
\usepackage{times}
\usepackage{graphicx}
\usepackage{placeins}  
\usepackage{setspace}
\usepackage{calc}
\usepackage{calrsfs}
\usepackage{float}
\usepackage{mathtools}
\usepackage{ulem}
\usepackage{amssymb,amsmath,amsfonts}
\usepackage{graphics,epsfig}
\usepackage{color,bm,hyperref}
\usepackage{ulem} 
\usepackage{xcolor}
\usepackage{tikz}
\usepackage{yfonts}
\usepackage{comment}
\usetikzlibrary{shapes.geometric,arrows}
\usetikzlibrary{decorations.markings}
\DeclareMathAlphabet{\pazocal}{OMS}{zplm}{m}{n}

\definecolor{lime}{HTML}{A6CE39}
\DeclareRobustCommand{\orcidicon}{\hspace{-4pt}
\begin{tikzpicture}
\draw[lime, fill=lime] (0,0)
circle [radius=0.16]
node[white] {\hspace{0.1mm}{\fontfamily{qag}\selectfont \tiny ID}};
\draw[white, fill=white] (-0.07,0.1)
circle [radius=0.01];
\end{tikzpicture}
\hspace{-3.2mm}
}
\foreach \x in {A, ..., Z}{\expandafter\xdef\csname
orcid\x\endcsname{\noexpand\href{https://orcid.org/\csname orcidauthor\x\endcsname}
{\noexpand\orcidicon}}
}

\begin{document}
\title{From Canonical to Tunable Phase Diagrams in Open Quantum Long-Range Systems}
\author{Anish Acharya\orcidA{} }
\email{Corresponding author; email: anish.acharya@tifr.res.in}
\affiliation{Department of Theoretical Physics, Tata Institute of Fundamental Research, Homi Bhabha Road, Mumbai 400005, India}
\author{Shamik Gupta\orcidC{} }
\affiliation{Department of Theoretical Physics, Tata Institute of Fundamental Research, Homi Bhabha Road, Mumbai 400005, India}

\begin{abstract}
{We investigate the dissipative dynamics of a generalized Lipkin-Meshkov-Glick (LMG) model coupled to a thermal environment. In this generalized model, in addition to the conventional quadratic interaction, one considers quartic interactions between spin-$1/2$'s coupled all-to-all and evolving in presence of a transverse field. Employing the usual linear Lindblad master equation with thermally-balanced jump processes, we derive magnetisation evolution equations, and demonstrate that the corresponding stationary solution reproduces the canonical equilibrium phase diagram of the model. We then extend our analysis to a nonlinear Lindblad equation that incorporates imperfect quantum-jump processes in terms of jump-retention parameters. Here, remarkably, the system relaxes to a genuine nonequilibrium stationary state whose properties differ qualitatively from those obtained in the linear case. The jump-retention parameters provide tunable knobs that shift the phase boundaries and even modify the nature of the phase transitions with respect to the linear case. Our exact results establish a direct connection between dissipative relaxation dynamics and stationary-state behavior, while identifying controlled quantum-jump retention as a mechanism for engineering nonequilibrium phases in long-range interacting open quantum systems.}
\end{abstract}
\maketitle

\section{Introduction}
Study of open quantum systems is central to contemporary quantum many-body physics studies, motivated both by fundamental questions that arise in the context of nonequilibrium quantum dynamics and by rapid experimental progress witnessed via engineered quantum platforms. In realistic settings, no quantum system is perfectly isolated, and interaction with the external environment results in dissipation, decoherence, and relaxation processes that can influence the collective behavior in a profound manner. The framework of quantum master equations, and in particular, that of the Gorini-Kossakowski-Sudarshan-Lindblad equation or simply the Lindblad equation~\cite{Lindblad:1976,Gorini:1976,Rivas:2012}, provides a powerful and widely-used framework to study Markovian open-system dynamics, enabling a systematic investigation of how equilibrium and nonequilibrium properties emerge in many-body open quantum systems~\cite{breuer2002,PhysRevLett.105.015702}. Physically, the Lindblad equation describes the evolution of an open quantum system undergoing continuous coherent evolution interrupted by stochastic quantum jumps induced by its environment, whose ensemble average yields a linear, completely positive, and trace-preserving evolution of the density matrix.

A central question concerns how an open quantum system approaches a stationary state under the combined influence of coherent many-body evolution and environmental dissipation. Over the years, studies have revealed that such dynamics may exhibit rich phenomenology, including control of relaxation timescales~\cite{4frd-ck2z,Koch_2016}, dissipative phase transitions~\cite{2fw8-bsjy,3wcr-sxtz}, and nontrivial dynamical scaling, all of which depend on the interplay between interactions and the nature of the coupling between the system and the environment modeled as a heat bath. Understanding these relaxation mechanisms is crucial for control and stabilization of quantum states in emerging quantum technologies~\cite{solanki2025, PhysRevB.108.054313}.

Long-range interacting quantum spin models offer a particularly fertile playground to explore relaxation dynamics in open quantum systems~\cite{RevModPhys.95.035002,King2025,Fiorelli_2023,Hryniuk2026,sulz2023open,PhysRevB.110.L081403}. Among them, the Lipkin-Meshkov-Glick (LMG) model is a paradigmatic example, which was originally proposed as a solvable many-body problem addressing atomic nuclei~\cite{LIPKIN1965188}, exhibits many interesting collective behavior and phase transitions. In the thermodynamic limit, the corresponding LMG Hamiltonian can be expressed in terms of collective spin operators, which enables the dynamics to be captured via a mean-field analysis. Recent studies have considered a generalized version of the LMG model, whereby, in addition to the conventional quadratic interaction, one incorporates a quartic four-spin interaction term. This extension substantially enriches the collective behaviour, leading to an equilibrium phase diagram exhibiting coexistence of continuous and first-order phase transitions separated by a tricritical point~\cite{PhysRevLett.133.050403,arrufatNicolo2025}.

Previous investigations of the generalized LMG model have focused on its equilibrium properties, deriving the phase diagram through a canonical partition-function approach~\cite{PhysRevLett.133.050403,arrufatNicolo2025}. While this analysis establishes the existence and nature of the equilibrium phases, an equally important question that remains to be addressed is: can one characterize analytically how such equilibrium states are dynamically attained when the system is coupled to an external heat bath? In particular, the relation between dissipative relaxation dynamics and the canonical stationary state remains largely unexplored for this model.

In this work, we address the aforesaid issue by studying the generalized LMG model in contact with a heat bath, within the framework of open quantum systems. Employing a Lindblad master equation with thermally-balanced excitation and relaxation processes, we derive rate equations in the form of closed mean-field evolution equations for the magnetization and analyze their long-time behavior. We demonstrate that the exact stationary solution of these rate equations reproduces the canonical equilibrium state and recovers the known equilibrium phase diagram of the model as reported in Refs.~\cite{PhysRevLett.133.050403,arrufatNicolo2025}. This establishes an explicit dynamical route through which the equilibrium phases emerge from dissipative evolution.

As emphasized earlier, a key feature of the standard Lindblad dynamics is the linearity of the master equation with respect to the density matrix. This linearity follows from the ensemble averaging of stochastic quantum trajectories, in which coherent evolution is punctuated by quantum jumps induced by the environment. Consequently, any modification of the underlying jump process can naturally give rise to nonlinear generalizations of the Lindblad equation. Recent theoretical developments have indeed unveiled situations in which quantum jumps are not perfectly realized, for example, due to either post-selection protocols, or, monitored dynamics, or, controlled retention of jump events. Such scenarios lead naturally to nonlinear generalizations of the Lindblad dynamics, whose studies have begun to attract considerable attention~\cite{PhysRevB.111.024303,PhysRevA.107.022216,kd3b-bfxq}. These nonlinear master equations open the possibility for the existence of qualitatively new stationary states and phase behavior.

Motivated by the aforementioned developments, we extend our analysis of the generalized LMG model beyond the conventional Lindblad framework by considering a nonlinear Lindblad evolution that incorporates imperfect quantum jumps. The resulting dynamics is governed by jump-retention parameters, which quantify the fraction of quantum jump events retained within the system. We then investigate the robustness of the canonical phase diagram, asking whether, and, if so, how, the retained jump events modify the phase boundaries and the nature of the associated phase transitions.

A central consequence of retaining quantum jumps is that the system always relaxes to a genuine nonequilibrium stationary state (NESS), which cannot be described by a canonical equilibrium distribution. This departure from equilibrium fundamentally alters the stationary properties of the system, giving rise to phase behavior that lies beyond that of both equilibrium and conventional linear Lindblad dynamics. In particular, the jump-retention parameters emerge as additional nonequilibrium control variables that shift the phase boundaries. Moreover, depending on whether these parameters are chosen symmetrically or asymmetrically, the nonlinear dynamics qualitatively changes the nature of the phase transitions, converting transitions that are continuous within the conventional Lindblad framework into first-order transitions, and vice versa. The resulting phase diagrams therefore exhibit a much richer structure than those obtained under equilibrium or standard dissipative conditions. 

In a related context, Ref.~\cite{PhysRevResearch.7.023057} demonstrated that optimizing system parameters can generate NESS in dissipative open quantum systems. By contrast, our work identifies controlled retention of quantum jumps as the physical mechanism responsible for the emergence of NESS and shows that it simultaneously enables systematic control over the phase boundaries and the nature of the associated phase transitions. These findings establish controlled dissipation as a powerful avenue for engineering collective phases in long-range quantum systems and provide new insights into the interplay between many-body interactions, bath-induced fluctuations, and nonlinear open-system dynamics.

The paper is organized as follows. We start with describing the generalised LMG model in Sec.~\ref{sec:sec1}. This is followed in Sec.~\ref{sec:Lindblad_LMG} by a discussion of the relaxation dynamics of the model attached to a heat bath and following the Lindblad evolution. Here, we present a detailed derivation of the magnetization rate equations in Sec.~\ref{sec2:subsecA}. We then discuss in Sec.~\ref{sec2:subsecC} the stationary state behavior and the resulting phase diagram. Next, we consider in Sec.~\ref{sec:sec3} the relaxation dynamics in presence of imperfect quantum jumps, and derive the corresponding stationary state in Sec.~\ref{sec3:subsec1}. We also discuss the modified phase diagram. We conclude the paper in Sec.~\ref{sec:conclu}. The two appendices contain further details of some calculations.

\section{Generalized LMG model: Canonical phase diagram}\label{sec:sec1}
We consider a long-range quantum spin system involving $N$ spin-$1/2$'s coupled all-to-all and interacting with Ising-like and four-spin interactions and in presence of a transverse field of strength $h$. The Hamiltonian of the system reads as~\cite{PhysRevLett.133.050403,arrufatNicolo2025}
\begin{align}
H=-\frac{J}{N}\left(\sum_{i=1}^N\sigma^{z}_i\right)^{2}-h\sum_{i=1}^N\sigma_i^{x}
-\frac{K}{N^{3}}\left(\sum_{i=1}^N\sigma_i^{z}\right)^{4}.
\label{eq:H}
\end{align}
Here, the operators $\sigma^{\mu}_i$, with $\mu=x,y,z$, are the usual Pauli matrices. 
We consider in this work the fully-ferromagnetic case with $J,K>0$. We will work in units in which $\hbar=1$.

With $K=0$, the Hamiltonian~\eqref{eq:H} becomes that of the celebrated Lipkin-Meshkov-Glick (LMG) model. This model possesses a $T=0$ quantum critical point at $h=h_{c}=J$, which corresponds to a phase transition between a paramagnetic phase with zero $z$-magnetization  and a ferromagnetic phase in which it has a non-zero value. With $K \ne 0$ and at any finite temperature, the phase transition between the paramagnetic and the ferromagnetic phase in the $(h/J,K/J)$-plane is either continuous or first-order depending on parameter regimes, with the two phase transition lines joining at a tricritical point. Our aim in this work is consider the system~\eqref{eq:H} in interaction with a heat bath in equilibrium at temperature $T$ and to derive, within a mean-field approximation, evolution equations in the form of rate equations for the quantum expectation of the components of the magnetization
\begin{align}
\mathbf{m}\equiv(m_x,m_y,m_z);~m_\mu \equiv \frac{1}{N}\sum_{i=1}^N\sigma_i^\mu,
\label{eq:mmu-defn}
\end{align}
by using a system-plus-reservoir approach. In this approach, assuming a time scale separation between the dynamics of the system and the heat bath, one integrates out the bath degrees of freedom to derive an effective dynamics for the system. The result is a Lindblad-like master equation for the density matrix or the density operator of the system~\cite{breuer2002}.  

\section{Relaxation dynamics via Lindblad evolution}\label{sec:Lindblad_LMG}
We model the dynamics of the system~\eqref{eq:H} in interaction with the heat bath via the Lindblad evolution for the density operator $\rho$ of the system, as
\begin{align}
    \frac{\partial \rho}{\partial t}=-i[H,\rho]+\mathcal{D}[\rho],
    \label{eq:Lindblad}
\end{align}
where we take the Lindblad dissipator $\mathcal{D}[\rho]$ to be of the form
\begin{align}
&\mathcal{D}[\rho]\nonumber \\
&=\sum_{\alpha=1,2}\gamma^{(\alpha)}\sum_{i=1}^N\Big(
L_i^{(\alpha)}\,\rho\,(L_i^{(\alpha)})^{\dagger}
-\frac{1}{2}\{(L_i^{(\alpha)})^\dagger L_i^{(\alpha)},\rho\}
\Big), \label{eq:Lindblad-1}
\end{align}
with the jump operators
\begin{align}
    L_i^{(1)}=\sigma_i^-,\quad L_i^{(2)}=\sigma_i^+.\label{eq:jump_operators}
\end{align}
Here, $\sigma^\pm=\sigma^x_i \pm i \sigma^y_i$ are the usual raising and lowering operators, respectively. Thus, the interaction with the bath  generates excitation and relaxation processes between
the eigenstates of $\sigma_i^z$. As will become clear from the form of the mean-field Hamiltonian in an appropriate basis derived below [see Eq.~\eqref{eq:HMF-2}], the ground state corresponds to the spin pointing up along $z$, while the excited state corresponds to the spin pointing down. Consequently, relaxation is effected by $\sigma_i^{+}$, whereas excitation is effected by $\sigma_i^{-}$. Considering the heat bath to be bosonic, the corresponding transition rates are~\cite{breuer2002}
\begin{align}
&\gamma^{(1)}=\gamma_0 n_{\mathrm{th}}, \nonumber \\
\label{eq:gamma-bosonic}\\
&\gamma^{(2)}=\gamma_0 (n_{\mathrm{th}}+1), \nonumber 
\end{align}
where
\begin{align}
n_{\mathrm{th}}
=\frac{1}{e^{\beta \Delta E}-1}
\end{align}
is the thermal expectation of the occupation number of the bath at inverse temperature
$\beta=1/(k_BT)$, and $\Delta E$ denotes the energy gap between the excited and the ground state of the system. We thus have
\begin{align}
     \mathcal{D}[\mathcal{O}] &=  \gamma^{(1)} \left( \sigma_i^- \mathcal{O} \sigma_i^+ - \frac{1}{2} \left\{\sigma_i^+ \sigma_i^-,\mathcal{O}\right\}\right) \nonumber\\
     &+\gamma^{(2)} \left( \sigma_i^+\mathcal{O} \sigma_i^- - \frac{1}{2} \left\{\sigma_i^- \sigma_i^+,\mathcal{O}\right\}\right).\label{eq:D_op-1}
\end{align}

In the thermodynamic limit $N \to \infty$,  in view of the all-to-all interaction in the Hamiltonian~\eqref{eq:H}, we may invoke the usual mean-field (MF) approximation, implying factorization of the total density operator into density operators for individual spins, $\rho =\otimes_{k=1}^N \rho_k$, with $\mathrm{Tr}[\rho_k]=1~\forall~k$~\cite{PhysRevLett.105.015702}. One then has $\rho_i =\mathrm{Tr}_{i'}[\rho]$, whereby one traces out all but the $i$-dependent variables. Under the MF approximation, Eq.~\eqref{eq:Lindblad} gives
\begin{align}
    \frac{\partial \rho_i}{\partial t}=-i\mathrm{Tr}_{i'}[H,\otimes_{k=1}^N\rho_k]+\mathcal{D}[\rho_i],
    \label{eq:Lindblad-1_MF}
\end{align}
wherein we have used $\mathrm{Tr}[\rho_k]=1~\forall~k$.

Let us evaluate the first term on the right-hand side (rhs) of Eq.~\eqref{eq:Lindblad-1_MF}. To this end, using Eq.~\eqref{eq:mmu-defn}, we get
\begin{align}
    \langle m_\mu \rangle=\mathrm{Tr}\left[\rho\frac{1}{N}\sum_{i=1}^N \sigma_i^\mu\right]&=\mathrm{Tr}\left[\left(\otimes_{k=1}^N\rho_k\right)\frac{1}{N}\sum_{i=1}^N \sigma_i^\mu\right]\nonumber \\
    &=\frac{1}{N}\sum_{i=1}^N \langle \sigma_i^\mu\rangle,
\end{align}
where we have defined $\langle \sigma_i^\mu \rangle \equiv \mathrm{Tr}[\sigma_i^\mu \rho_i]$. Since the system~\eqref{eq:H} is homogeneous, the expectation $\langle \sigma_i^\mu \rangle$ takes up the same value for all $i$, resulting in
\begin{align}
    \langle m_\mu \rangle=\langle \sigma_i^\mu \rangle.
    \label{eq:m-mu}
\end{align}
Next, consider the operator $\mathcal{O}_1 \equiv \sum_{l,m=1}^N \sigma_l^\mu \sigma_m^\mu$. By singling out the $i$-th contribution, one gets 
\begin{align}
\mathrm{Tr}_{i'}\left[\mathcal{O}_1\otimes_{k=1}^N\rho_k\right] 
&= (\sigma_i^\mu)^2 \rho_i + \sigma_i^\mu \rho_i \left( \sum_{m \ne i} \langle \sigma_m^\mu \rangle \right) \nonumber\\
&+ \left( \sum_{l \ne i} \langle \sigma_l^\mu \rangle \right) \sigma_i^\mu \rho_i + \rho_i \sum_{l,m \ne i} \langle \sigma_l^\mu \rangle \langle\sigma_m^\mu \rangle.
\end{align}
We want to now evaluate the term $\mathrm{Tr}_{i'}[\mathcal{O}_1,\rho]$. Defining $\chi \equiv \sum_{l\neq i} \sigma^\mu_l$ and $\mathbb{B} \equiv \sum_{l,m \ne i} \langle \sigma_l^\mu \rangle \langle \sigma_m^\mu \rangle$, we have
\begin{align}
&\mathrm{Tr}_{i'}[\mathcal{O}_1,\rho]\nonumber \\
&= [(\sigma_i^{\mu})^2, \rho_i ]+\langle \chi \rangle[\sigma_i^\mu, \rho_i]+\langle \chi \rangle[\sigma_i^\mu, \rho_i]+[\rho_i,\mathbb{B}]\nonumber\\
    &=2 \langle \chi \rangle [\sigma_i^\mu, \rho_i],\label{eq:Trace_op1}
\end{align}
where we have $\langle \chi \rangle= \sum_{l \ne i} \langle \sigma_l^\mu \rangle=N \langle m_\mu \rangle - \langle \sigma_i^\mu \rangle$, and we have utilized the fact that $(\sigma_i^{\mu})^2 = \mathbb{I}$ for spin-$1/2$ operators, causing the first commutator on the rhs in the second step to vanish. Furthermore, since $\mathbb{B}$ is an expectation value, the term $[\rho_i,\mathbb{B}]$ vanishes. Then, we can write for the operator $\mathrm{O}_1=-(J/N)\mathcal{O}_1$ with $\mu=z$, which corresponds to the first term of the Hamiltonian~\eqref{eq:H}, that
\begin{align}
     \mathrm{Tr}_{i'}[\mathrm{O}_1,\rho]&=-2 \frac{J}{N}\left(N \langle m_z \rangle - \langle \sigma_i^z \rangle \right)[\sigma_i^z, \rho_i]\nonumber\\
     &=-[2 J \langle m_z \rangle \sigma_i^z, \rho_i],\label{eq:Tr_O1_exp1}
\end{align}
where, in the second step, we have taken the limit $N \to \infty$. 

Proceeding similarly for the operator $\mathcal{O}_2=(\sum_j \sigma_j^\mu)^4= (\sigma_i^\mu + \sum_{j \neq i} \sigma_j^\mu)^4$, we get 
\begin{align}
\mathrm{Tr}_{i'}\left[\mathcal{O}_2\otimes_{k=1}^N\rho_k\right] 
&= (\sigma_i^\mu)^4\rho_i + 4 \sigma_i^\mu \rho_i\langle\chi^3\rangle+ 4\langle \chi \rangle \sigma_i^\mu \rho_i \nonumber\\
&+ 6\rho_i \langle \chi^2 \rangle+\rho_i \langle \chi^4\rangle,\label{eq:Tr_iprime_exp1}
\end{align}
where we have $\langle \chi^m \rangle_{m \ge 1}= \mathrm{Tr}_{i'} \left[(\otimes_{k \neq i} \rho_k) \left(\sum_{j \neq i} \sigma_j^\mu \right)^m \right]$.
Equation~\eqref{eq:Tr_iprime_exp1} may be used to write $\mathrm{Tr}_{i'}[\mathcal{O}_2,\rho]$ as
\begin{align}
    \mathrm{Tr}_{i'}[\mathcal{O}_2,\rho]= 4\langle\chi^3\rangle[\sigma_i^\mu , \rho_i]+ 4\langle\chi\rangle[\sigma_i^\mu , \rho_i].
\end{align}
Now, for the operator $\mathrm{O}_2 \equiv -(K/N^3)\mathcal{O}_2$ with $\mu=z$, corresponding to the quartic term in the Hamiltonian~\eqref{eq:H}, we get
\begin{align}
&\mathrm{Tr}_{i'}[\mathrm{O}_2,\rho] \nonumber \\
&= -\frac{4K}{N^3}\left(\langle\chi^3\rangle + \langle\chi\rangle\right)[\sigma_i^z, \rho_i] \nonumber \\
&= -\frac{4K}{N^3}\left( \sum_{j\neq i,k \neq i,l\neq i} \langle \sigma_j^z \sigma_k^z \sigma_l^z \rangle^3 + (N \langle m_z \rangle - \langle \sigma_i^z \rangle) \right)\nonumber\\
&\times[\sigma_i^z, \rho_i],\label{eq:Tr_O2_exp1}
\end{align}
where we have $\langle \sigma_j^z \sigma_k^z \sigma_l^z \rangle=\mathrm{Tr}_{jkl}[\rho_j \rho_k \rho_l \sigma_j^z \sigma_k^z \sigma_l^z ]$.
To evaluate the first term on the rhs of the above equation carefully, we separately consider the contributions of the indices $(j,k,l)$, to get
\begin{align}
     \sum_{j\neq i,k\neq i,l\neq i} \langle \sigma_j^z \sigma_k^z \sigma_l^z \rangle ^3 &= \mathcal{C}_1\sum_{j=k=l} \langle \sigma_j^z \rangle+ 3~ \mathcal{C}_2\sum_{j=l\neq k} \langle \sigma_k^z \rangle\nonumber\\
     &+ \mathcal{C}_3\sum_{j\neq k\neq l} \langle \sigma_j^z \rangle \langle \sigma_k^z \rangle \langle \sigma_l^z \rangle,\label{eq:combinations}
\end{align}
where on the rhs, we have $j\neq i$, $k \neq i$, $l \neq i$, along with $\mathcal{C}_1= (N-1)$, $\mathcal{C}_2= (N-1)(N-2)$, and $\mathcal{C}_3= (N-1)(N-2)(N-3)$. Using Eq.~\eqref{eq:combinations} in Eq.~\eqref{eq:Tr_O2_exp1}, and then taking the limit $N\to \infty$, we obtain the leading-order contribution as
\begin{align}
\mathrm{Tr}_{i'}[\mathrm{O}_2,\rho] 
&= -[4K\langle m_z\rangle^3\sigma_i^z, \rho_i].\label{eq:Tr_O2_exp2}
\end{align}
Similarly, for the operator $\mathrm{O}_3=-h \mathcal{O}_3;~\mathcal{O}_3 \equiv \sigma^x_i+\sum_{j\neq i}\sigma_j^x$, corresponding to the field term in the Hamiltonian~\eqref{eq:H}, we may show that
\begin{align}
    \mathrm{Tr}_{i'}[\mathrm{O}_3,\rho] 
&=-[h \sigma_i^x, \rho_i]. \label{eq:Tr_O3_exp1}
\end{align}

Using Eqs.~\eqref{eq:Tr_O1_exp1},~\eqref{eq:Tr_O2_exp2},~\eqref{eq:Tr_O3_exp1} in Eq.~\eqref{eq:Lindblad-1_MF}, we obtain
\begin{align}
    \frac{\partial \rho_i}{\partial t}=-i[H_\mathrm{MF},\rho_i]+\mathcal{D}[\rho_i],
    \label{eq:MF-Lindblad}
\end{align}
where 
\begin{align}
H_\mathrm{MF}=-\mathbb{H}\sigma^z_i - h \sigma^x_i;~\mathbb{H}\equiv 2J \langle m_z \rangle + 4K \langle m_z \rangle^3
\label{Eq:H_MF_Final}
\end{align}
is the MF Hamiltonian. 

\subsection{Magnetization rate equations}\label{sec2:subsecA}
We start with Eq.~\eqref{eq:m-mu}. Differentiating both sides of the  equation with respect to time, and noting that only $\rho$ is time-dependent, we get on using Eq.~\eqref{eq:MF-Lindblad} that
\begin{align}
    \frac{d \langle m_\mu \rangle}{dt}&=-i\mathrm{Tr}\left[ [H_\mathrm{MF}, \rho_i]\sigma_i^\mu\right]+ \mathrm{Tr}\left[\mathcal{D}[ \rho_i]\sigma_i^\mu \right]\nonumber\\
    &=-i \mathrm{Tr}\left[ \rho_i [\sigma_i^\mu, H_\mathrm{MF}]\right]+ \mathrm{Tr}\left[\rho_i \mathcal{D}^\dagger[\sigma_i^\mu]\right],\label{eq:m_eq2}
\end{align}
where in the last line, we have used $\mathrm{Tr}[[A,B]C]=\mathrm{Tr}[B[C,A]]$ and have used the identity $\mathrm{Tr}\left[\mathcal{D}[ \rho_i]\sigma_i^\mu \right]= \mathrm{Tr}\left[\rho_i \mathcal{D}^\dagger[\sigma_i^\mu]\right]$ that may be checked by direct substitution. Here, we have 
\begin{align}
\mathcal{D}^\dagger[\mathcal{O}]&=\gamma^{(1)} \left( \sigma_i^+ \mathcal{O} \sigma_i^- - \frac{1}{2} \left\{\sigma_i^+ \sigma_i^-,\mathcal{O}\right\}\right)\nonumber \\
     &+\gamma^{(2)} \left( \sigma_i^- \mathcal{O} \sigma_i^+ - \frac{1}{2} \left\{\sigma_i^- \sigma_i^+,\mathcal{O}\right\}\right).\label{eq:D_op_dagger}
\end{align}

Now, the mean-field Hamiltonian~\eqref{Eq:H_MF_Final} may be rewritten by using the effective magnetic field $\mathbb{G} \equiv \sqrt{\mathbb{H}^2+h^2}$ as
\begin{align}
    &H_\mathrm{MF}=-\mathbb{G}\left(\frac{\mathbb{H}}{\mathbb{G}}\sigma_i^z+\frac{h}{\mathbb{G}} \sigma_i^x\right).
\end{align}
Let us define an angle $\theta$ such that  $\cos{\theta}\equiv\mathbb{H}/\mathbb{G}$ and $\sin{\theta}\equiv h/\mathbb{G}$, which allows us to define a new basis with modified Pauli operators as
\begin{align}
    &\sigma_i^{z'}= \sigma_i^z \cos{\theta} +\sigma_i^x \sin{\theta},\nonumber\\
    &\sigma_i^{x'}= \sigma_i^x \cos{\theta} -\sigma_i^z \sin{\theta},\nonumber\\
    &\sigma_i^{y'}=\sigma_i^y.\nonumber\\
    &\label{eq:basis_mu'}
\end{align}
Consequently, one has the mean-field Hamiltonian in the transformed basis as
\begin{align}
    &H_\mathrm{MF}'=-\mathbb{G}\, (\sigma_i^z)'.
    \label{eq:HMF-2}
\end{align}
In this new basis, Eq.~\eqref{eq:MF-Lindblad} reads as
\begin{align}
    \frac{\partial \rho_i'}{\partial t}=-i[H_\mathrm{MF}',\rho_i']+\mathcal{D}'[\rho_i'],
    \label{eq:MF-Lindblad-1}
\end{align}
with
\begin{align}
     \mathcal{D}'[\mathcal{O}] &=  \gamma^{(1)} \left( (\sigma_i^-)' \mathcal{O} (\sigma_i^+)' - \frac{1}{2} \left\{(\sigma_i^+)' (\sigma_i^-)',\mathcal{O}\right\}\right) \nonumber\\
     &+\gamma^{(2)} \left( (\sigma_i^+)'\mathcal{O} (\sigma_i^-)' - \frac{1}{2} \left\{(\sigma_i^-)' (\sigma_i^+)',\mathcal{O}\right\}\right),\label{eq:D_op-11}
\end{align}
where $\gamma^{(1)}$ and $\gamma^{(2)}$ are given by Eq.~\eqref{eq:gamma-bosonic} with $\Delta E=2\mathbb{G}$. 
The stationary state of Eq.~\eqref{eq:MF-Lindblad-1} is the canonical-equilibrium state 
$\rho_{i,\mathrm{eq}}'\propto e^{-\beta H_\mathrm{MF}'}=e^{\beta \mathbb{G}\,(\sigma_i^z)'}$. Indeed, we have $[H_\mathrm{MF}',\rho_{i,\mathrm{eq}}']=0$. On the other hand, using the identity $e^{X \sigma^z}=  \mathbb{I}~\cosh{X} + \sigma^z~\sinh{X}$, we get $e^{\beta \mathbb{G}~(\sigma_i^z)'}= \mathbb{I} ~\cosh{\beta \mathbb{G}}+ (\sigma_i^z)'~\sinh{\beta \mathbb{G}}$, yielding
\begin{align}
    \mathcal{D}'[e^{\beta \mathbb{G} (\sigma_i^z)'}] &= -\gamma^{(1)} (\sigma_i^z)'\left(\cosh{\beta \mathbb{G}}+\sinh{\beta \mathbb{G}}\right)\nonumber\\
    &+\gamma^{(2)} (\sigma_i^z)' \left(\cosh{\beta \mathbb{G}}-\sinh{\beta \mathbb{G}}\right)\nonumber\\
    &=\left(-\gamma^{(1)} e^{\beta \mathbb{G}}+\gamma^{(2)} e^{-\beta \mathbb{G}}\right)~(\sigma_i^z)'\nonumber \\
    &=0,\label{Eq:D_linear_pth}
\end{align}
where we have used Eq.~\eqref{eq:gamma-bosonic} with $\Delta E= 2\mathbb{G}$. Using Eq.~\eqref{eq:basis_mu'}, we then obtain the canonical-equilibrium stationary state of the dynamics~\eqref{eq:MF-Lindblad} as $\rho_{i,\mathrm{eq}} \propto e^{\beta\mathbb{G}(\sigma_i^z \cos{\theta} +\sigma_i^x \sin{\theta})}=e^{\beta (\mathbb{H}\sigma_i^z+h\sigma_i^x )}=e^{-\beta H_\mathrm{MF}}$.

In the primed basis, Eq.~\eqref{eq:m_eq2} reads as
\begin{align}
    \frac{d \langle m_\mu' \rangle}{dt}&=-i \mathrm{Tr}\left[ \rho_i' [(\sigma_i^\mu)', H_\mathrm{MF}']\right]+ \mathrm{Tr}\left[\rho_i' (\mathcal{D}')^\dagger [(\sigma_i^\mu)']\right].\label{eq:m_eq2-1}
\end{align}
On putting $\mu=x$ in Eq.~\eqref{eq:m_eq2-1}, we get that
\begin{align}
    \frac{d \langle m_x'\rangle}{dt}&=2 \mathbb{G} \langle m_y'\rangle+ \mathrm{Tr}\left[\rho_i'(\mathcal{D}')^\dagger[(\sigma_i^x)'] \right],\label{eq:mx_eq1}
\end{align}
where we have used $[(\sigma^x_i)', (\sigma^z_i)']= -2 i (\sigma^y_i)'$. We now simplify the second term on the rhs of  Eq.~\eqref{eq:mx_eq1}. 
Using the identities $(\sigma_i^x)'=(\sigma_i^+)' +(\sigma_i^-)'$, $(\sigma_i^y)'=-i((\sigma_i^+)'-(\sigma_i^-)')$, $(\sigma_i^+)' (\sigma_i^-)'= (1/2)(\mathbb{I}+(\sigma_i^z)')$, and $(\sigma_i^{+})^{'2}=(\sigma_i^{-})^{'2}=0$, one obtains
\begin{align}
    (\mathcal{D}')^\dagger[(\sigma_i^x)']=-\frac{1}{2}(\gamma^{(1)}+\gamma^{(2)})(\sigma_i^x)', 
\end{align}
yielding $\mathrm{Tr}\left[\rho_i' (\mathcal{D}')^\dagger[(\sigma_i^x)'] \right]= - (1/2)(\gamma^{(1)}+\gamma^{(2)})\langle m_x' \rangle$, so that we finally have from Eq.~\eqref{eq:mx_eq1} that
\begin{align}
    \frac{d \langle m_x'\rangle}{dt}=2 \mathbb{G} \langle m_y'\rangle- \frac{1}{2}(\gamma^{(1)}+\gamma^{(2)})\langle m_x'\rangle.
\end{align}
Proceeding similarly, and on the basis of the result that $\mathrm{Tr}\left[\rho_i' (\mathcal{D}')^\dagger[(\sigma_i^y)'] \right]=-(1/2)(\gamma^{(1)}+\gamma^{(2)})\langle m_y'\rangle$, we obtain
\begin{align}
    \frac{d \langle m_y'\rangle }{dt}=-2 \mathbb{G} \langle m_x'\rangle- \frac{1}{2}(\gamma^{(1)}+\gamma^{(2)})\langle m_y'\rangle.
\end{align}
For $\mu=z$, the first term in Eq.~\eqref{eq:m_eq2-1} vanishes, while the second term is obtained as 
\begin{align}
\mathrm{Tr}\!\left[\rho_i' (\mathcal{D}')^\dagger[(\sigma_i^z)'] \right]\!\!=\!-(\gamma^{(1)}+\gamma^{(2)})\! \left(\!\langle m_z' \rangle\!-\frac{\gamma^{(2)}-\gamma^{(1)}}{\gamma^{(1)}+\gamma^{(2)}}\!\right),\label{eq:D_dagger_sigmaz_eta0}
\end{align}
so that
\begin{align}
    \frac{d \langle m_z'\rangle }{dt}=-(\gamma^{(1)}+\gamma^{(2)})\!\!\left(\!\langle m_z' \rangle\!-\!\frac{\gamma^{(2)}-\gamma^{(1)}}{\gamma^{(1)}+\gamma^{(2)}}\!\right).\label{eq:dmz/dt_eqv1}
\end{align}

Let us define $\lambda$ as $\lambda \equiv \gamma^{(1)}+\gamma^{(2)}$. On the other hand, with $\Delta E= 2 \mathbb{G}$, we have on using Eq.~\eqref{eq:gamma-bosonic} that
\begin{align}
    \frac{\gamma^{(2)}-\gamma^{(1)}}{\gamma^{(1)}+\gamma^{(2)}}= \tanh{\beta \mathbb{G}}.\label{eq:gamma_ratio_bosonic}
\end{align}
Equation~\eqref{eq:dmz/dt_eqv1} then gives
\begin{align}
     \frac{d \langle m_z'\rangle}{dt}&=-\lambda \left(\langle m_z'\rangle -\tanh{\beta \mathbb{G}}\right).
\end{align}  
We have therefore the magnetization rate equations as
\begin{align}
& \frac{d \langle m_x'\rangle}{dt}=2 \mathbb{G} \langle m_y'\rangle- \frac{\lambda}{2} \langle m_x'\rangle, \nonumber \\
&\frac{d \langle m_y'\rangle}{dt}=-2 \mathbb{G} \langle m_x'\rangle- \frac{\lambda}{2} \langle m_y'\rangle,\nonumber\\
    &\frac{d \langle m_z'\rangle}{dt}=-\lambda \left(\langle m_z'\rangle -\tanh{\beta \mathbb{G}}\right).\nonumber\\ \label{eq:rate_eqs_for_Gamma_zero}
 \end{align}
 
If the bath is taken to be fermionic, one has $n_\mathrm{th}= 1/(e^{\beta  \Delta E}+1)$. Then, we have 
\begin{align}
   & \frac{\gamma^{(2)}-\gamma^{(1)}}{\gamma^{(1)}+\gamma^{(2)}}
    =1 - 2 n_\mathrm{th}= 1- \frac{2}{e^{\beta  \Delta \mathbb{H}}+1}= \tanh{\beta \mathbb{G}},
\end{align}
which is identical to Eq.~\eqref{eq:gamma_ratio_bosonic}. We thus see that the rate equations for the magnetization remain unchanged, irrespective of whether the bath is bosonic or fermionic.

We want to now write the rate equations~\eqref{eq:rate_eqs_for_Gamma_zero} in the unprimed system. To this end, inverting Eq.~\eqref{eq:basis_mu'}, we get
\begin{align}
    &\sigma_i^{z}= \sigma_i^{z'} \cos{\theta} -\sigma_i^{x'} \sin{\theta},\nonumber\\
    &\sigma_i^{x}= \sigma_i^{x'} \cos{\theta} +\sigma_i^{z'} \sin{\theta},\nonumber\\
    &\sigma_i^{y}=\sigma_i^{y'},\nonumber\\
    &\label{eq:basis_mu}
\end{align}
and these equations also hold for the transformation between the primed and the unprimed magnetization components, $\langle m_x'\rangle$ and $\langle m_x\rangle$, etc. Using them, along with Eq.~\eqref{eq:rate_eqs_for_Gamma_zero}, we finally obtain
\begin{align}
    &\frac{d \langle m_z\rangle }{dt}=\!-2 h \langle m_y\rangle\! -\!\lambda  \frac{\mathbb{H}^2}{\mathbb{G}^2} \langle m_z\rangle\!-\!\frac{\lambda}{2}\frac{h^2}{\mathbb{G}^2} \langle m_z\rangle\!-\!\frac{\lambda}{2}\frac{\mathbb{H}h}{\mathbb{G}^2}\langle m_x\rangle\nonumber \\
    &+\frac{\lambda \mathbb{H}}{\mathbb{G}}\tanh{(\beta \mathbb{G})},\nonumber\\
    &\frac{d \langle m_y\rangle}{dt}=- 2 \mathbb{H} \langle m_x\rangle+2 h \langle m_z\rangle -\frac{\lambda}{2}\langle m_y\rangle,\nonumber\\
    &\frac{d \langle m_x\rangle}{dt}= 2 \mathbb{H} \langle m_y\rangle -\lambda \langle m_x\rangle \frac{h^2}{\mathbb{G}^2}-\frac{\lambda}{2}\langle m_x\rangle \frac{\mathbb{H}^2}{\mathbb{G}^2}\!-\!\frac{\lambda}{2}\frac{\mathbb{H}h}{\mathbb{G}^2}\langle m_z\rangle\nonumber \\
    &+\lambda\frac{h}{\mathbb{G}}\tanh{(\beta \mathbb{G})},\nonumber\\\label{eq:rate_equations_Lindblad}
\end{align}
which may be compactly written in terms of $\mathbf{m}\equiv (m_x,m_y,m_z)$ as
\begin{align}
\frac{d{\langle \mathbf{m}\rangle}}{dt}
&=\begin{pmatrix}
-\frac{\lambda\left(\mathbb{H}^2+2h^2\right)}{2\mathbb{G}^2}
& 2\mathbb{H}
& -\frac{\lambda \mathbb{H} h}{2\mathbb{G}^2}
\\
-2\mathbb{H}
& -\frac{\lambda}{2}
& 2h
\\
-\frac{\lambda \mathbb{H} h}{2\mathbb{G}^2}
& -2h
& -\frac{\lambda\left(2\mathbb{H}^2+h^2\right)}{2\mathbb{G}^2}
\end{pmatrix}
\langle \mathbf{m}\rangle\nonumber \\
&+
\lambda \tanh(\beta \mathbb{G})
\begin{pmatrix}
\frac{h}{\mathbb{G}}\\
0\\
\frac{\mathbb{H}}{\mathbb{G}}
\end{pmatrix}.
\label{eq:compact_evolution_Gamma_not_zero}
\end{align}

\subsection{The stationary state: Recovering the canonical phase diagram}\label{sec2:subsecC}
The stationary-state solution of Eq.~\eqref{eq:compact_evolution_Gamma_not_zero} can be easily obtained as
\begin{align}
    &\langle m_x\rangle_\mathrm{eq}=\frac{h}{\mathbb{G}}\tanh{(\beta \mathbb{G})},\nonumber \\
    &\langle m_y\rangle_\mathrm{eq}=0,\nonumber \\
    &\langle m_z\rangle_\mathrm{eq}=\frac{\mathbb{H}}{\mathbb{G}}\tanh{(\beta \mathbb{G})}.\nonumber \\\label{eq:stationary_state_LMG}
\end{align}
Putting $\mathbb{G}=\sqrt{\mathbb{H}^2+h^2}$ and $\mathbb{H}=2 J \langle m_z\rangle_\mathrm{eq}+ 4K \langle m_z\rangle_\mathrm{eq}^3$, we obtain in particular the stationary-state $z$-magnetization to be given by the self-consistent equation
\begin{align}
    &\langle m_z\rangle_\mathrm{eq}=\mathcal{G}(\langle m_z\rangle_\mathrm{eq});\label{eq:mz_eq_LME}\\ &\mathcal{G}(\langle m_z\rangle_\mathrm{eq})\equiv \frac{2 J \langle m_z\rangle_\mathrm{eq} + 4 K \langle m_z\rangle_\mathrm{eq}^3}{\sqrt{h^2+(2 J \langle m_z\rangle_\mathrm{eq} + 4 K \langle m_z\rangle_\mathrm{eq}^3)^2}} \nonumber \\
    &\times\tanh\;\left({\beta \sqrt{h^2+(2 J \langle m_z\rangle_\mathrm{eq} + 4 K \langle m_z\rangle_\mathrm{eq}^3)^2}}\right). \label{eq:mz_stationarystate}
\end{align}
Note that $\mathcal{G}(m)$ is an odd function of $m$.

Now, according to the principles of equilibrium statistical mechanics, the stationary-state magnetization is obtained as the minimizer of the Landau free energy $\mathcal{F}$ of the system, which, consistent with Eq.~\eqref{eq:mz_stationarystate}, is given by
\begin{align}
    \mathcal{F}(\langle m_z\rangle)=F_0+\int_0^{\langle m_z\rangle}dy~\left(y-\mathcal{G}(y)\right),\label{eq:landau_free_energy1}
\end{align}
where $F_0$ is a constant. The following expansion follows:
\begin{align}
    \mathcal{F}(\langle m_z\rangle)&\!=\!\!F_0 + \frac{a}{2} \langle m_z\rangle^2 + \frac{b}{4} \langle m_z\rangle^4+\frac{c}{6} \langle m_z\rangle^6\nonumber\\
    &+\mathcal{O}(\langle m_z\rangle^8),\label{eq:landau_free_energy2_LME}
\end{align}
with the coefficients $a$, $b$ and $c$ given by
\begin{align}
&a\equiv \left(1 - \frac {2 J\tanh(\beta h)}{h}\right),\\
&b\equiv
\frac{4}{h^3}
\left(
-h^2K\tanh(\beta h)
-h\beta J^3\operatorname{sech}^2(\beta h)\right.\nonumber\\
&\left.\qquad\quad
+J^3\tanh(\beta h)
\right),\nonumber\\
&c\equiv
\frac{J^2\operatorname{sech}^2(\beta h)}{h^5}
\Bigl[6\left(J^3-2h^2K\right)
\left(2h\beta-\sinh(2\beta h)\right)
\nonumber\\
&\qquad\quad+8h^2J^3\beta^2\tanh(\beta h)
\Bigr].\label{eq:a_b_c}
\end{align}
Note that the above coefficients depend on $J, h, K$ and $\beta$. 

Following Ref.~\cite{PhysRevLett.133.050403}, we now study phase transitions in $\langle m_z\rangle_\mathrm{eq}$ by fixing $\beta$ and $J$ such that $\beta J=2/3$ and varying $K$ and $h$. In particular, we may ask how at a fixed $K$ the behavior of $\langle m_z\rangle_\mathrm{eq}$ changes on varying $h$. To this end, invoking the Landau theory of phase transitions, we may truncate $\mathcal{F}(\langle m_z \rangle)$ from Eq.~\eqref{eq:landau_free_energy2_LME} at $\mathcal{O}(\langle m_z \rangle^4)$, provided $b>0$. In such a case, one has a continuous phase transition between a zero and a non-zero value of $\langle m_z\rangle_\mathrm{eq}$, signalled by $a$ becoming zero on tuning the parameter $h$ at a fixed $K$. In the $(h,K)$-plane, one then obtains the continuous transition line as given by the equation 
\begin{align}
    h = 2 J \tanh{(\beta h)}.
    \label{eq:continuous}
\end{align}
\begin{figure}[htbp!]
\centering
\includegraphics[width=6 cm,height=5 cm]{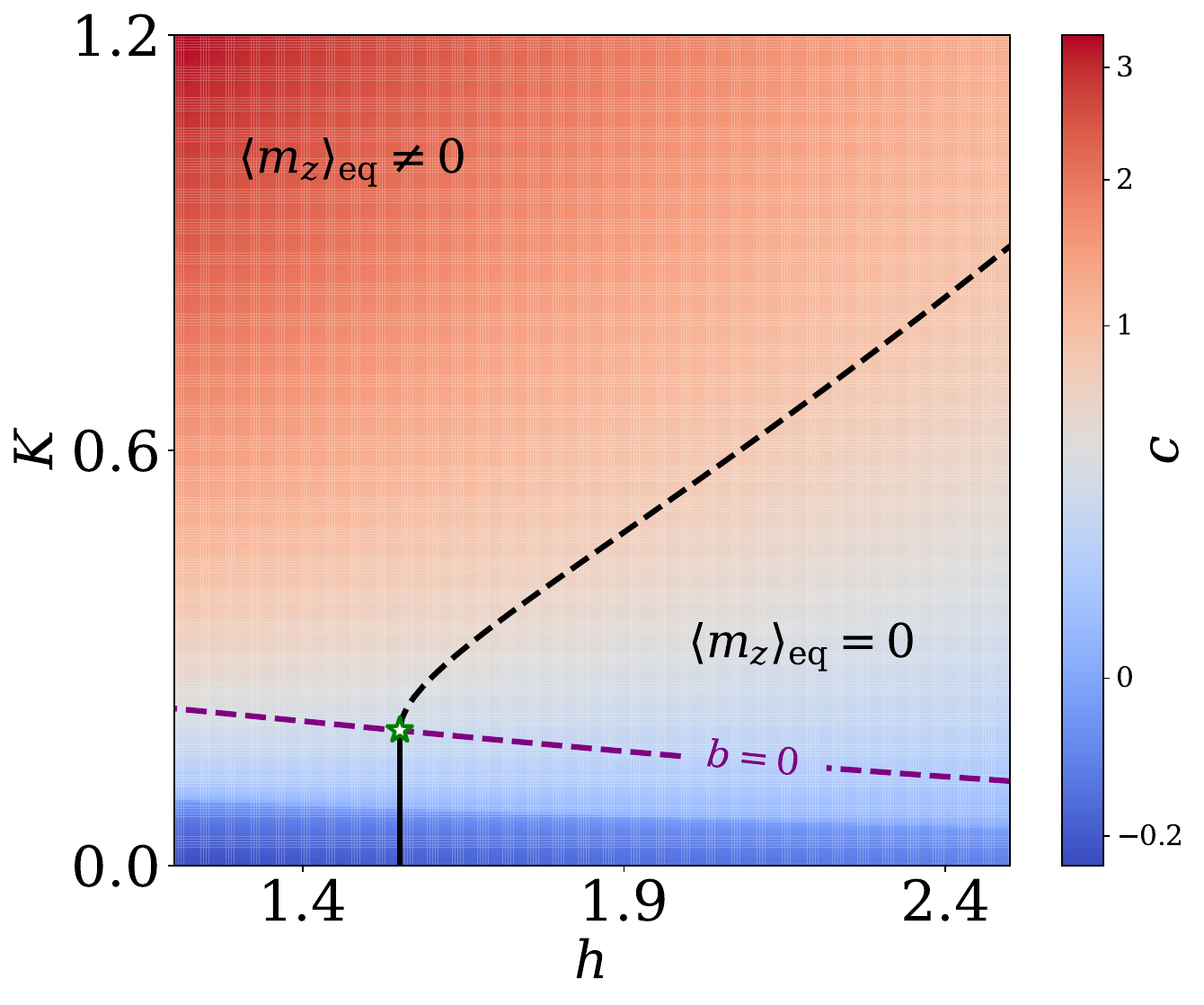}
\caption{Canonical phase diagram of the model~\eqref{eq:H} in the $(h,K)$-plane, obtained as the stationary state solution of the Lindblad equation~\eqref{eq:Lindblad} and at $\beta J=2/3$, along with behavior of the Landau coefficients $b$ and $c$. We also show the continuous transition line~\eqref{eq:continuous} (black solid line) and the first-order transition line (black dashed line). The color map shows the sixth-order coefficient $c$, while the violet (dashed) contour denotes the locus $b=0$, passing through the tricritical point~\eqref{eq:tcp} (green star). Above the contour, we have $b<0$, while one has $b>0$ below it. In the region with $b<0$, one has $c>0$. }
\label{fig:Fig1}
\end{figure}
Setting $a=b=0$, one obtains the condition for the existence of a tricritical point that marks the endpoint of the continuous transition line~\eqref{eq:continuous}. The equation for the tricritical point is obtained as
\begin{align}
    & h= 2 J \tanh{(\beta h)},\\
    &K=\frac {J^3} {h^2} + \frac {\beta J^2} {2} \left (1 - \frac {4  J^2} {h^2} \right).
    \label{eq:tcp}
\end{align}
On the other hand, when $b < 0$, thermodynamic stability requires truncating at $\mathcal{O}(\langle m_z \rangle^6)$ term in Eq.~\eqref{eq:landau_free_energy2_LME}, provided the corresponding coefficient $c$ is positive. As verified in Fig.~\ref{fig:Fig1}, indeed, the coefficient $c$ remains strictly positive in the $(h,K)$-plane in the region with $b<0$, allowing us to safely effect the said truncation. In such a case, one may obtain the first-order transition line by requiring coexistence of the disordered phase and the ordered phase. Numerically, this is determined by locating the parameter values for which the minimizers of $\mathcal{F}$, occurring at a zero value and two non-zero values $\pm m^*$, with $m^*>0$, are such that $\mathcal{F}(\langle m^* \rangle) = \mathcal{F}(0)$.

Following the above procedure, we show in Fig.~\ref{fig:Fig1} and for $\beta J=2/3$ the continuous transition line~\eqref{eq:continuous}, the tricritical point~\eqref{eq:tcp}, and the first-order transition line. We have hereby reproduced the canonical phase diagram of the Hamiltonian~\eqref{eq:H} reported in Ref.~\cite{PhysRevLett.133.050403,arrufatNicolo2025}, through a dynamical setting within the Lindblad equation approach.

\section{Lindblad evolution with imperfect quantum jumps}\label{sec:sec3}
Our analysis of the dynamics of an open quantum system presented thus far can be extended to incorporate the case of post-selected quantum dynamics, where a fraction of jump events is systematically retained within the system. Specifically, for the generalized LMG model~\eqref{eq:H}, we will consider the non-linear Lindblad equation (NLME)~\cite{PhysRevB.111.024303,PhysRevA.107.022216}
\begin{widetext}
 \begin{align}
\frac{d \rho_i}{d t}= -i [H_\mathrm{MF}, \rho_i]+\mathcal{D}[\rho_i];~\mathcal{D}[\rho_i]=\sum_{\alpha}\gamma^{(\alpha)}\left((1-\eta_\alpha)  L^{(\alpha)}_i \rho_i L^{(\alpha)\dagger}_i-\frac{1}{2}\{L^{(\alpha)\dagger}_i L^{(\alpha)}_i, \rho_i \}+ \eta_\alpha \langle L^{(\alpha)\dagger}_i L^{(\alpha)}_i\rangle \rho_i\right), \label{eq:NLME_1}
\end{align}
\end{widetext}
where we have $\langle L^{(\alpha)\dagger} L^{(\alpha)} \rangle = \mathrm{Tr}[L^{(\alpha)\dagger} L^{(\alpha)} \rho_i]$ (see details of the derivation of the above equation in Appendix~\ref{sec:app1}). Note that the term due to unitary evolution is the same in the above equation and in Eq.~\eqref{eq:MF-Lindblad}. The parameter $\eta_\alpha \in [0,1]$ represents the retention probability of the quantum jump along channel $\alpha$. To allow for full generality, we consider distinct retention probabilities, $\eta_1$ and $\eta_2$, corresponding to the jump operators $L_i^{(1)}$ and $L_i^{(2)}$, respectively. One may note that the structure of Eq.~\eqref{eq:NLME_1} guarantees for any $\eta_\alpha$ that the trace of the density matrix is preserved at all times. Setting $\eta_1=\eta_2=0$ reduces Eq.~\eqref{eq:NLME_1} to the Lindblad equation~\eqref{eq:MF-Lindblad}. For the case at hand, one has in place of Eq.~\eqref{eq:D_op_dagger} the expression 
\begin{align}
    &\mathcal{D}^\dagger[\mathcal{O}]\nonumber\\
    &=\gamma^{(1)} \left( (1-\eta_1)\sigma^+_i \mathcal{O} \sigma^-_i - \frac{1}{2} \{\sigma^+_i \sigma^-_i, \mathcal{O}\} + \eta_1 \langle \sigma^+_i \sigma^-_i \rangle \mathcal{O} \right)\nonumber \\
    &+\gamma^{(2)} \left((1-\eta_2)\sigma^-_i \mathcal{O} \sigma^+_i - \frac{1}{2} \{\sigma^-_i \sigma^+_i, \mathcal{O}\} + \eta_2 \langle \sigma^-_i \sigma^+_i \rangle \mathcal{O} \right).\nonumber\\
    \label{eq:D_dagger_NLME_v2}
\end{align}

For Eq.~\eqref{eq:NLME_1}, written in the primed basis, the stationary state is not the canonical-equilibrium state 
$\rho_{i,\mathrm{eq}}'\propto e^{-\beta H_\mathrm{MF}'};~H_\mathrm{MF}'=-\mathbb{G}(\sigma_i^z)'$, unlike the case for Eq.~\eqref{eq:MF-Lindblad-1}. To check whether $\rho_{i,\mathrm{eq}}'$ is the stationary-state solution of Eq.~\eqref{eq:NLME_1}, we proceed as follows. As before, we have $[H_\mathrm{MF}',\rho_{i,\mathrm{eq}}']=0$. On the other hand, we have $\mathcal{D}^{'}[\rho_i']=\mathcal{D}^{'}_{\mathrm{s}}[\rho_i'] + \mathcal{D}^{'}_{\mathrm{nl}}[\rho_i']$, where $\mathcal{D}^{'}_{\mathrm{s}}$ refers to the dissipator term as in the linear Lindblad equation, Eq.~\eqref{eq:MF-Lindblad-1}, which we have already seen to satisfy $\mathcal{D}^{'}_{\mathrm{s}}[\rho_{i,\mathrm{eq}}'] = 0$ in Sec.~\ref{sec2:subsecA}. The term $\mathcal{D}^{'}_{\mathrm{nl}}[\rho_{i,\mathrm{eq}}']$ is given by
\begin{align}
    \mathcal{D}^{'}_{\mathrm{nl}}[\rho_{i,\mathrm{eq}}'] = \sum_{\alpha=1}^2 \gamma^{(\alpha)} \eta_\alpha \left( \langle L^{(\alpha)\dagger}_i L^{(\alpha)}_i \rangle \rho_i - L^{(\alpha)}_i \rho_i L^{(\alpha)\dagger}_i \right).\label{eq:D_prime_nl_1}
\end{align}
Writing $\rho_{i,\mathrm{eq}}'=a_+ (\sigma_i^+)' (\sigma_i^-)' + a_- (\sigma_i^-)' (\sigma_i^+)'$, with $a_\pm= e^{\pm \beta \mathbb{G}}/ Z$ and $Z=e^{\beta \mathbb{G}}+e^{-\beta \mathbb{G}}$, one may show
$\langle (\sigma_i^+)' (\sigma_i^-)'\rangle = a_+$ and $\langle (\sigma_i^-)' (\sigma_i^+)'\rangle = a_-$, and that
\begin{align}
    \mathcal{D}^{'}_{\mathrm{nl}} \left[ \rho_{i,\mathrm{eq}}'\right]
    &=  \left( \gamma^{(1)}\eta_1 a_+^2 -\gamma^{(2)}\eta_2 a_-^2\right) \!(\sigma_i^z)',\label{eq:D_nl}
\end{align}
on using $(\sigma_i^+)' (\sigma_i^-)'- (\sigma_i^-)' (\sigma_i^+)'=(\sigma_i^z)'$. It is then evident that for an arbitrary choice of the retention probabilities $\eta_1$ and $\eta_2$, one has $\mathcal{D}^{'}_{\mathrm{nl}} \left[ \rho_{i,\mathrm{eq}}'\right] \neq 0$, which proves that $\rho_{i,\mathrm{eq}}'$ is not a stationary state of Eq.~\eqref{eq:NLME_1} in the primed basis. The resulting stationary state is therefore a genuine nonequilibrium stationary state (NESS).

To proceed, let us note here some useful identities:
\begin{align}
    &\sigma^-_i\sigma^z_i\sigma^+_i=\frac{1}{2}(\mathbb{I}-\sigma^z_i)~~,~~\sigma^+_i\sigma^z_i\sigma^-_i=-\frac{1}{2}(\mathbb{I}+\sigma^z_i),\nonumber\\
    &\{\sigma^+_i\sigma^-_i,\sigma^z_i \}=(\mathbb{I}+\sigma^z_i)~~,~~\{\sigma^-_i\sigma^+_i,\sigma^z_i \}=(\sigma^z_i-\mathbb{I});
\end{align}
these identities hold also for the primed basis. 
We first evaluate Eq.~\eqref{eq:D_dagger_NLME_v2} in the primed basis and for $\mathcal{O}=(\sigma^z_i)'$; the term multiplying $\gamma^{(2)}$ may be evaluated as follows:
\begin{align}
    &(1-\eta_2)(\sigma^-_i)' (\sigma^z_i)' (\sigma^+_i)' - \frac{1}{2} \{(\sigma^-_i)' (\sigma^+_i)', (\sigma^z_i)'\} \nonumber \\
    &+ \eta_2 \langle (\sigma^-_i)' (\sigma^+_i)' \rangle (\sigma^z_i)'\nonumber\\
    &=\frac{2-\eta_2}{2}\mathbb{I}+\frac{1}{2}\left(2 \eta_2 - 2 -\eta_2 \langle m_z' \rangle\right)(\sigma^z_i)';
\end{align}
One may in a similar manner simplify the term multiplying $\gamma^{(1)}$, to finally obtain
\begin{align}
    (\mathcal{D}')^\dagger[(\sigma^z_i)']
     &= \gamma^{(2)}\left(1-\frac{\eta_2}{2}\right)\mathbb{I} -  \gamma^{(1)}\left(1-\frac{\eta_1}{2}\right)\mathbb{I}\nonumber\\
     &+\gamma^{(2)} \left(\eta_2-1-\frac{\eta_2 \langle m_z'\rangle}{2} \right)(\sigma^z_i)'\nonumber\\
     &-\gamma^{(1)}\left( 1-\eta_1-\frac{\eta_1 \langle m_z'\rangle}{2}\right)(\sigma^z_i)'.
\end{align}
Using $\gamma^{(2)}+\gamma^{(1)}=\lambda$ and $\gamma^{(2)}-\gamma^{(1)}=\lambda~\delta$, with $\delta\equiv \tanh(\beta \mathbb{G})$, see Eq.~\eqref{eq:gamma_ratio_bosonic}, and defining
\begin{align}
    \Gamma\equiv \frac{(1+\delta)}{2}\left(1- \frac{\eta_2}{2}\right)-\frac{(1-\delta)}{2} \left(1-\frac{\eta_1}{2}\right),\label{eq:Gamma_def}
\end{align}
and 
\begin{align}
    \Sigma (\langle m_z'\rangle) &\equiv (1+\delta)\left(\eta_2-1-\frac{\eta_2 \langle m_z'\rangle}{2}\right)\nonumber\\
    &+(1-\delta)\left(\eta_1-1+\frac{\eta_1 \langle m_z'\rangle}{2} \right),\label{eq:Sigma_def}
\end{align}
one obtains
\begin{align}
    (\mathcal{D}')^\dagger[(\sigma^z_i)']
    &= \lambda~\Gamma \mathbb{I}+\frac{\lambda}{2} \Sigma(\langle m_z'\rangle)~(\sigma^z_i)'.\label{eq:D_dagger_sigmaz_i_1}
\end{align}
Following similar steps, one may obtain 
\begin{align}
    &(\mathcal{D}')^\dagger[(\sigma^x_i)']=-\frac{\lambda}{2}\Lambda(\langle m_z\rangle) (\sigma^x_i)', \nonumber \\
    &(\mathcal{D}')^\dagger[(\sigma^y_i)'] = -\frac{\lambda}{2}\Lambda(\langle m_z'\rangle)(\sigma^y_i)', \nonumber \\
\end{align}
where we have defined 
\begin{align}
&\Lambda(\langle m_z'\rangle)\equiv \nonumber \\
&1-\frac{1}{2} \left(\eta_2 (1+\delta) (1-\langle m_z'\rangle)+\eta_1 (1-\delta) (1-\langle m_z'\rangle)\right).
\end{align}

Reverting to the unprimed basis, we get 
\begin{align}
&\mathcal{D}^\dagger[\sigma^z_i]\nonumber\\ 
&=\frac{\lambda}{2} \Sigma(\langle m_{z}\rangle)~\sigma^z_i \cos^2\theta + \frac{\lambda}{2} \sin\theta \cos\theta\left(\Sigma(\langle m_{z}\rangle)\right.\nonumber\\
&~~\left.+\Lambda(\langle m_{z} \rangle)\right)\sigma^x_i-\frac{\lambda}{2} \Lambda(\langle m_{z}\rangle)~\sigma^z_i \sin^2\theta + \lambda \cos\theta~\Gamma \mathbb{I},\label{eq:D_dagg_z_NLME_unprime}
\end{align}
and similarly, one has 
\begin{align}
&\mathcal{D}^\dagger[\sigma^x_i]\nonumber\\
&=\frac{\lambda}{2} \Sigma(\langle m_{z}\rangle)~\sigma^x_i \sin^2\theta + \frac{\lambda}{2} \sin\theta \cos\theta\left(\Sigma(\langle m_{z}\rangle)\right.\nonumber\\
&~~\left.+\Lambda(\langle m_{z} \rangle)\right)\sigma^z_i
-\frac{\lambda}{2} \Lambda(\langle m_{z}\rangle)~\sigma^x_i \cos^2\theta + \lambda \sin\theta~\Gamma \mathbb{I}, \nonumber \\
&\mathcal{D}^\dagger[\sigma^y_i]=-\frac{\lambda}{2}\Lambda(\langle m_{z} \rangle)\sigma^{y},\nonumber \\ \label{eq:D_dagg_y_NLME_unprime}
\end{align}
with
\begin{align}
     &\Sigma (\langle m_{z}\rangle) = (1+\delta)\left(\eta_2-1-\frac{\eta_2}{2} \frac{\langle m_z\rangle \mathbb{H} +\langle m_x\rangle h}{\mathbb{G}}\right)\nonumber\\
     &+(1-\delta)\left(\eta_1-1+\frac{\eta_1}{2} \frac{\langle m_z\rangle \mathbb{H} +\langle m_x\rangle h}{\mathbb{G}} \right),\label{eq:sigma_labframe}
\end{align}
and
\begin{align}
&\Lambda(\langle m_z\rangle)\equiv 1-\frac{1}{2} \left(\eta_2 (1+\delta) \left(1-\frac{\langle m_z\rangle \mathbb{H} +\langle m_x\rangle h}{\mathbb{G}}\right)\right.\nonumber\\
&\left.+\eta_1 (1-\delta) \left(1-\frac{\langle m_z\rangle \mathbb{H} +\langle m_x\rangle h}{\mathbb{G}}\right)\right).\label{eq:lambda_labframe}
\end{align}
Using $\cos\theta=\mathbb{H}/\mathbb{G}$ and $\sin\theta=h/\mathbb{G}$, we finally get
\begin{align}
&\mathcal{D}^\dagger[\sigma^z_i]=\frac{\lambda}{2} \Sigma \frac{\mathbb{H}^2}{\mathbb{G}^2}~\sigma^z_i + \frac{\lambda}{2} \frac{h \mathbb{H}}{\mathbb{G}^2}\left(\Sigma+\Lambda\right)\sigma^x_i\nonumber\\
&\qquad\qquad-\frac{\lambda}{2} \Lambda\frac{h^2}{\mathbb{G}^2}~\sigma^z_i + \lambda \frac{\mathbb{H}}{\mathbb{G}}~\Gamma \mathbb{I},\nonumber\\
&\mathcal{D}^\dagger[\sigma^x_i]=\frac{\lambda}{2} \Sigma \frac{h^2}{\mathbb{G}^2}~\sigma^x_i + \frac{\lambda}{2} \frac{h \mathbb{H}}{\mathbb{G}^2}\left(\Sigma+\Lambda\right)\sigma^z_i\nonumber\\
&\qquad\qquad-\frac{\lambda}{2} \Lambda\frac{\mathbb{H}^2}{\mathbb{G}^2} \sigma^x_i + \lambda\frac{h}{\mathbb{G}}~\Gamma \mathbb{I},\nonumber\\
&\mathcal{D}^\dagger[\sigma^y_i]=-\frac{\lambda}{2}\Lambda~\sigma^{y}_i, \nonumber \\ \label{eq:dissipator_terms_NLME}
\end{align}
where we have dropped the argument of $\Sigma$ and $\Lambda$ for the sake of brevity. 

\subsection{Magnetization rate equations}
Using the results presented thus far, and proceeding as was done to obtain Eq.~\eqref{eq:rate_equations_Lindblad}, one finally obtains corresponding to Eq.~\eqref{eq:NLME_1} the magnetization rate equations as
\begin{align}
\frac{d \langle m_z\rangle}{dt}
&=-2h~\langle m_y \rangle +\langle m_z \rangle \frac{\lambda(\mathbb{H}^2 \Sigma-h^2 \Lambda)}{2 \mathbb{G}^2}\nonumber\\
&+\langle m_x \rangle\frac{\lambda \mathbb{H}h}{2 \mathbb{G}^2}
(\Sigma+\Lambda)+\lambda\frac{\mathbb{H}}{\mathbb{G}}\Gamma, \nonumber \\
\frac{d \langle m_y\rangle}{dt}
&=-2\mathbb{H}\langle m_x \rangle+ 2h \langle m_z\rangle-\frac{\lambda}{2}\langle m_y \rangle \Lambda, \nonumber \\
\frac{d \langle m_x \rangle}{dt}&=2\mathbb{H}\langle m_y \rangle +\langle m_x \rangle \frac{\lambda (h^2\Sigma-\mathbb{H}^2\Lambda)}{2 \mathbb{G}^2}\nonumber\\
&+\langle m_z \rangle \frac{\lambda \mathbb{H}h}{2\mathbb{G}^2}
(\Sigma+\Lambda) +\lambda \frac{h}{\mathbb{G}}
\Gamma,\nonumber \\ \label{eq:dmxdt_NLME}
\end{align}
which may be written in a compact form as 
\begin{align}
\frac{d \langle \mathbf m \rangle}{dt}
&=\begin{pmatrix} \frac{\lambda(h^2 \Sigma - \mathbb{H}^2 \Lambda)}{2\mathbb{G}^2} & 2\mathbb{H} & \frac{\lambda \mathbb{H} h (\Sigma + \Lambda)}{2\mathbb{G}^2} \\ -2\mathbb{H} & -\frac{\lambda}{2} \Lambda & 2h \\ \frac{\lambda \mathbb{H} h (\Sigma + \Lambda)}{2\mathbb{G}^2} & -2h & \frac{\lambda(\mathbb{H}^2 \Sigma - h^2 \Lambda)}{2\mathbb{G}^2} \end{pmatrix}
\langle \mathbf m \rangle\nonumber\\
&+\lambda \Gamma
\begin{pmatrix}
\dfrac{h}{\mathbb{G}}\\
0\\
\dfrac{\mathbb{H}}{\mathbb{G}}
\end{pmatrix}.\label{eq:compact_evolution_h_eta}
\end{align}

\subsection{The stationary state and the phase diagram}\label{sec3:subsec1}
Equation~\eqref{eq:compact_evolution_h_eta} yields the stationary state 
\begin{align}
& \langle m_x \rangle_{\mathrm{ss}}=-\frac{2~h~\Gamma}{\mathbb{G}~\Sigma}, \nonumber \\
& \langle m_y \rangle_{\mathrm{ss}}= 0, \nonumber\\
& \langle m_z \rangle_{\mathrm{ss}}=-\frac{2~\mathbb{H}~ \Gamma}{\mathbb{G}~\Sigma}, \nonumber\\\label{eq:stationary_state_NLME}
\end{align}
where we have $\Sigma=\Sigma (\langle m_z\rangle_\mathrm{ss})$ and $\Gamma=\Gamma (\langle m_z\rangle_\mathrm{ss})$.
Setting $\eta_1=0=\eta_2$ correctly reproduces the canonical equilibrium result, Eq.~\eqref{eq:stationary_state_LMG}. Note that unlike Eq.~\eqref{eq:stationary_state_LMG}, the stationary state (ss) here is not a canonical equilibrium state, but is an NESS.

As in Sec.~\ref{sec2:subsecC}, we focus here on the $z$-magnetization obtained in the NESS. Rewriting Eq.~\eqref{eq:stationary_state_NLME} by defining $\mathrm{M} \equiv - 2 \Gamma / \Sigma$, we get 
\begin{align}
    \langle m_z \rangle_\mathrm{ss}  =\frac{\mathbb{H}}{\mathbb{G}}\mathrm{M};~~\langle m_x  \rangle_\mathrm{ss} =\frac{h}{\mathbb{G}}\mathrm{M}.\label{eq:mz_mx_m_Representation}
\end{align}
Note that $\mathrm{M}$ implicitly depends on $\beta$ through $\Gamma$ and $\Sigma$ via $\delta$, see Eqs. \eqref{eq:Gamma_def} and \eqref{eq:Sigma_def}. Putting  $\langle m_z \rangle_\mathrm{ss}$ and $\langle m_x  \rangle_\mathrm{ss}$ from Eq.~\eqref{eq:mz_mx_m_Representation} in Eq.~\eqref{eq:sigma_labframe}, we get
\begin{align}
\Sigma&= (1+\delta)(\eta_2-1)+(1-\delta) (\eta_1-1)\nonumber\\
&+ \mathrm{M}\left(\frac{\eta_1}{2}(1-\delta)-\frac{\eta_2}{2}(1+\delta) \right).\label{eq:Sigma_defined_with_m}
\end{align}
Defining $M_0 \equiv (1+\delta)(\eta_2-1)+(1-\delta) (\eta_1-1)$ and $A \equiv \left((\eta_1/2)(1-\delta)-(\eta_2/2)(1+\delta) \right)$, we obtain from Eq.~\eqref{eq:Sigma_defined_with_m} and on using $\Sigma=-2\Gamma/\mathrm{M}$ the following quadratic equation for $\mathrm{M}$:
\begin{align}
    \mathrm{M}^2 A +\mathrm{M} M_0+ 2 \Gamma=0.\label{eq:m_eq_quadratic}
\end{align}
Out of the $2$ possible roots of Eq.~\eqref{eq:m_eq_quadratic}, which are
\begin{align}
    \mathrm{M}_\pm=\frac{-M_0\pm \sqrt{M_0^2-8 A \Gamma}}{2 A},\label{eq:two_roots}
\end{align}
only one represents a physically-meaningful solution, as will be clear in the following discussion. To this end, let us note that setting $\eta_\alpha =0~(\alpha=1,2)$ yields results for the usual Lindblad dynamics discussed in Sec.~\ref{sec:Lindblad_LMG}. Out of the two solutions in Eq.~\eqref{eq:two_roots}, we identify the physically-meaningful one by analyzing their behavior as $\eta_\alpha\to 0$, when we have $\Gamma=\delta~,M_0=-2~,A= 0$. In this limit, one gets $\mathrm{M}_+=2/A - \Gamma$, which clearly diverges for $A=0$. On the other hand, in the same limit, we get $\mathrm{M}_-= \delta$, which on using in Eq.~\eqref{eq:mz_mx_m_Representation} yields the Lindblad stationary-state result, Eq.~\eqref{eq:stationary_state_LMG}. Henceforth, we will consider 
\begin{align}
\mathrm{M}_-=\frac{-M_0-\sqrt{M_0^2-8 A \Gamma}}{2 A},\label{eq:correct_root_m}
\end{align}
as the physical solution of Eq.~\eqref{eq:m_eq_quadratic}. The stationary-state $z$-magnetization is now given by the self-consistent relation
\begin{align}
    \langle m_z \rangle_\mathrm{ss}= \mathcal{G}(\langle m_z \rangle_\mathrm{ss})=\frac{\mathbb{H}~\mathrm{M}_- }{\mathbb{G}}.\label{eq:mz_eq_NLME}
\end{align}

Now, unlike the procedure following Eq.~\eqref{eq:mz_eq_LME}, we need to adopt a different strategy to characterize the phase transition of $\langle m_z \rangle_\mathrm{ss}$, since the resulting stationary state is an NESS, as discussed following Eq.~\eqref{eq:D_nl}. To proceed, we refer to Appendix~\ref{sec:app2}, and write Eq.~\eqref{eq:mz_eq_NLME} as
\begin{align}
   &\langle m_z\rangle_\mathrm{ss}=\mathcal{G}(\langle m_z\rangle_\mathrm{ss});\nonumber \\
   &\mathcal{G}(\langle m_z\rangle)\!\equiv\!\mathcal{A} \langle m_z\rangle\! +\! \mathcal{B} \langle m_z\rangle^3\!+\! \mathcal{C} \langle m_z\rangle^5
    \!+\!\mathcal{O}(\langle m_z\rangle^7),\label{eq:cal_G_NLME}
\end{align}
with the coefficients $\mathcal{A}$,  $\mathcal{B}$ and $\mathcal{C}$ given by
\begin{align}
    &\mathcal{A}\equiv \frac{2 J \mathrm{M}^{(0)}}{h},\nonumber\\
    & \mathcal{B}\equiv -\frac{4 J^3 \mathrm{M}^{(0)}}{h^3}+\frac{4 K \mathrm{M}^{(0)}}{h}+\frac{2 J \mathrm{M}^{(2)}}{h},\nonumber\\
    & \mathcal{C}\equiv \!  \frac{2J \mathrm{M}^{(4)}}{h} - \frac{4(J^3 - K h^2)\mathrm{M}^{(2)}}{h^3}\nonumber\\
   &\qquad\quad+\frac{12(J^5 - 2 K J^2 h^2) \mathrm{M}^{(0)}}{h^5}, \nonumber \\ \label{eq:expression_cal_A_B_C}
\end{align}
where we have defined $\mathrm{M}^{(0)}\equiv \mathrm{M}_- (\langle m_z \rangle=0)$ and $\mathrm{M}^{(2k)}\equiv (1/(2k)!)\partial^{2k} \mathrm{M}_- / \partial \langle m_z \rangle^{2k}|_{\langle m_z \rangle=0}; k=1,2$. While deriving the series expression in Eq.~\eqref{eq:cal_G_NLME}, we exploit the fact that $\mathrm{M}_-(\langle m_z \rangle)$ from Eq.~\eqref{eq:correct_root_m} is an even function of $\langle m_z \rangle$. This stems from the effective magnetic field $\mathbb{G}$ being an even function of $ \langle m_z \rangle$. Since all constituent parameters of $\mathrm{M}_-$, including $M_0$, $\Gamma$ and $A$, depend on $\mathbb{G}$ via $\delta$ (recall that $\delta= \tanh({\beta \mathbb{G}}))$, the quantity $\mathrm{M}_-(\langle m_z \rangle)$ becomes an even function. Hence, $\mathrm{M}_-$ may be written as $\mathrm{M}_-= \mathrm{M}^{(0)} +\sum_{k=1}^\infty \mathrm{M}^{(2k)} \langle m_z \rangle^{2k}$, finally resulting in Eq.~\eqref{eq:cal_G_NLME} as an odd-powered series of $\langle m_z \rangle$. Here, we also note that following Eq.~\eqref{eq:expression_cal_A_B_C}, the coefficients $\mathcal{A}, \mathcal{B}$ and $\mathcal{C}$ depend on all  the system parameters $J, h, K, \beta$ and the retention probabilities $\eta_1, \eta_2$.

\begin{figure}[h]
\centering
\includegraphics[width=\linewidth, height=0.44\linewidth]{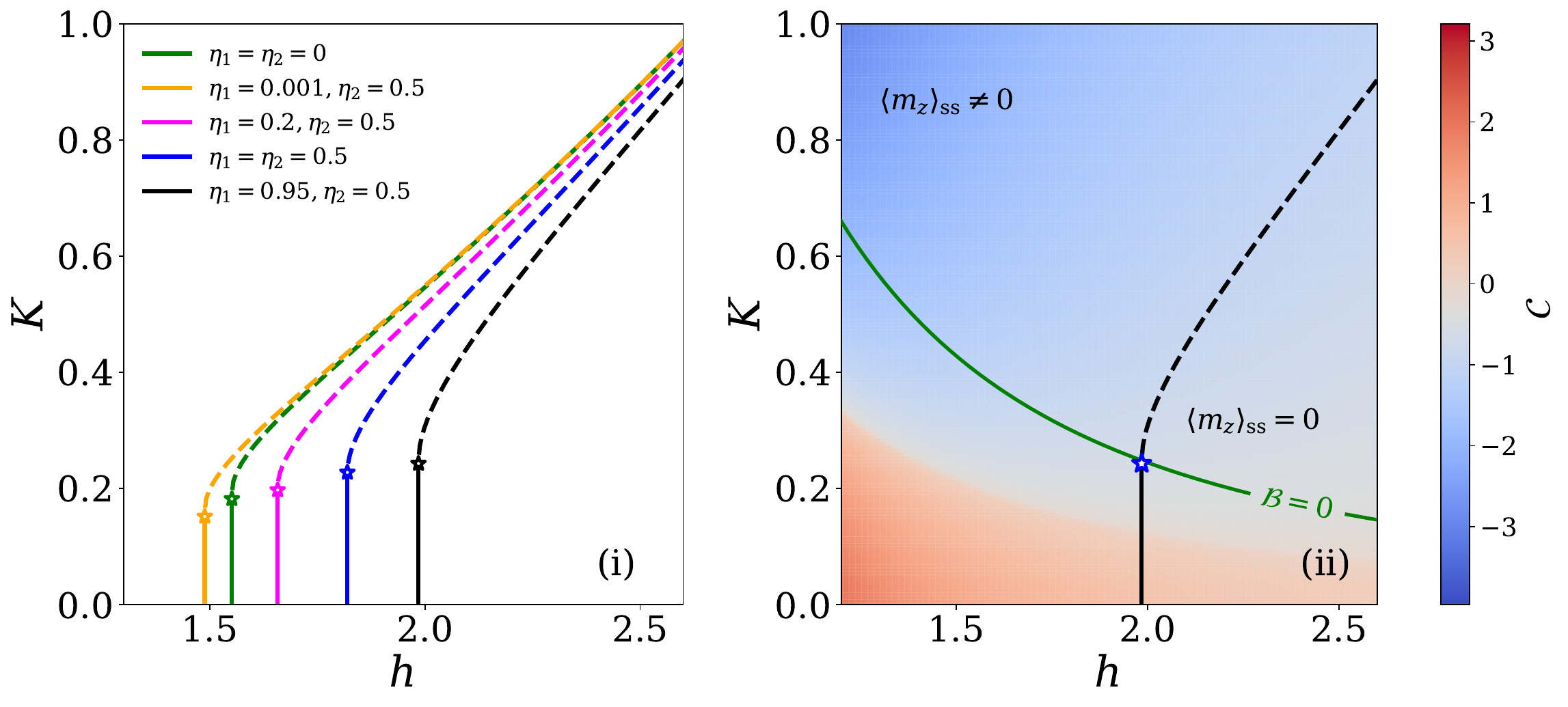}
\caption{Phase diagram of the model~\eqref{eq:H} in the $(h,K)$-plane, obtained as the NESS solution of the NLME~\eqref{eq:NLME_1} for different choices of the retention probabilities $\eta_1$ and $\eta_2$ and at $\beta J =2/3$ (panel ($\mathrm{i}$)). The solid lines denote the continuous transition line~\eqref{eq:continuous_NLME}, the star denotes the corresponding tricritical point, while dashed lines denote the first-order transition line. Panel ($\mathrm{ii}$) shows for $\eta_1=0.95,~\eta_2=0.5$ the color map of the coefficient $\mathcal{C}$ in Eq.~\eqref{eq:expression_cal_A_B_C}, while the green (dashed) contour denotes the locus of $\mathcal{B}=0$, passing through the corresponding tricritical point (blue star). Above the contour, we have $\mathcal{B}>0$, while below it, one has $\mathcal{B}<0$. In the region with $\mathcal{B}>0$, one has $\mathcal{C}<0$.}
\label{fig:NLME_kh_diagram}
\end{figure}

As was done in Sec.~\ref{sec:Lindblad_LMG}, we now study the phase transition in $\langle m_z \rangle_{\rm{ss}}$ by fixing $\beta$ and $J$ such that $\beta J=2/3$. Here we may ask how at fixed values of $K$, $\eta_1$ and $\eta_2$ the behavior of $\langle m_z \rangle_{\rm{ss}}$ changes upon varying $h$. Following the criterion detailed in Appendix~\ref{sec:app2}, one gets a continuous phase transition in $\langle m_z \rangle$ when the conditions $\mathcal{A}=1$, $\mathcal{B}<0$ are satisfied upon tuning the parameter $h$ at fixed values of $K, \eta_1$ and $\eta_2$. In the $(h,k)$-plane, one then obtains the condition for continuous transition as
\begin{align}
    h= 2 J \mathrm{M}^{(0)}.
    \label{eq:continuous_NLME}
\end{align}
Setting $\mathcal{A}=1$ and $\mathcal{B}=0$, one obtains the condition for the existence of a tricritical point that marks the endpoint of the continuous transition line~\eqref{eq:continuous_NLME}. For $\mathcal{B}>0$, provided $\mathcal{C}<0$, one may obtain the first-order transition line by finding a value $\mathcal{A}=\mathcal{A}^*$ such that Eq.~\eqref{eq:cal_G_NLME} has a zero and a non-zero solution $\langle m_z^*\rangle$ coexisting. As verified in Fig.~\ref{fig:NLME_kh_diagram}$(\rm{ii})$ for a representative choice of $\eta_1, \eta_2$, the coefficient $\mathcal{C}$ remains negative in the $(h,K)$-plane in the region with $\mathcal{B}>0$. We show in Fig.~\ref{fig:NLME_kh_diagram}$(\rm{i})$ how the continuous transition line~\eqref{eq:continuous_NLME} and the first-order transition line get modified depending on different choices of the probability parameters $\eta_1$ and $\eta_2$.

\begin{figure}[h]
\centering
\includegraphics[width=0.6\linewidth, height=0.6\linewidth]{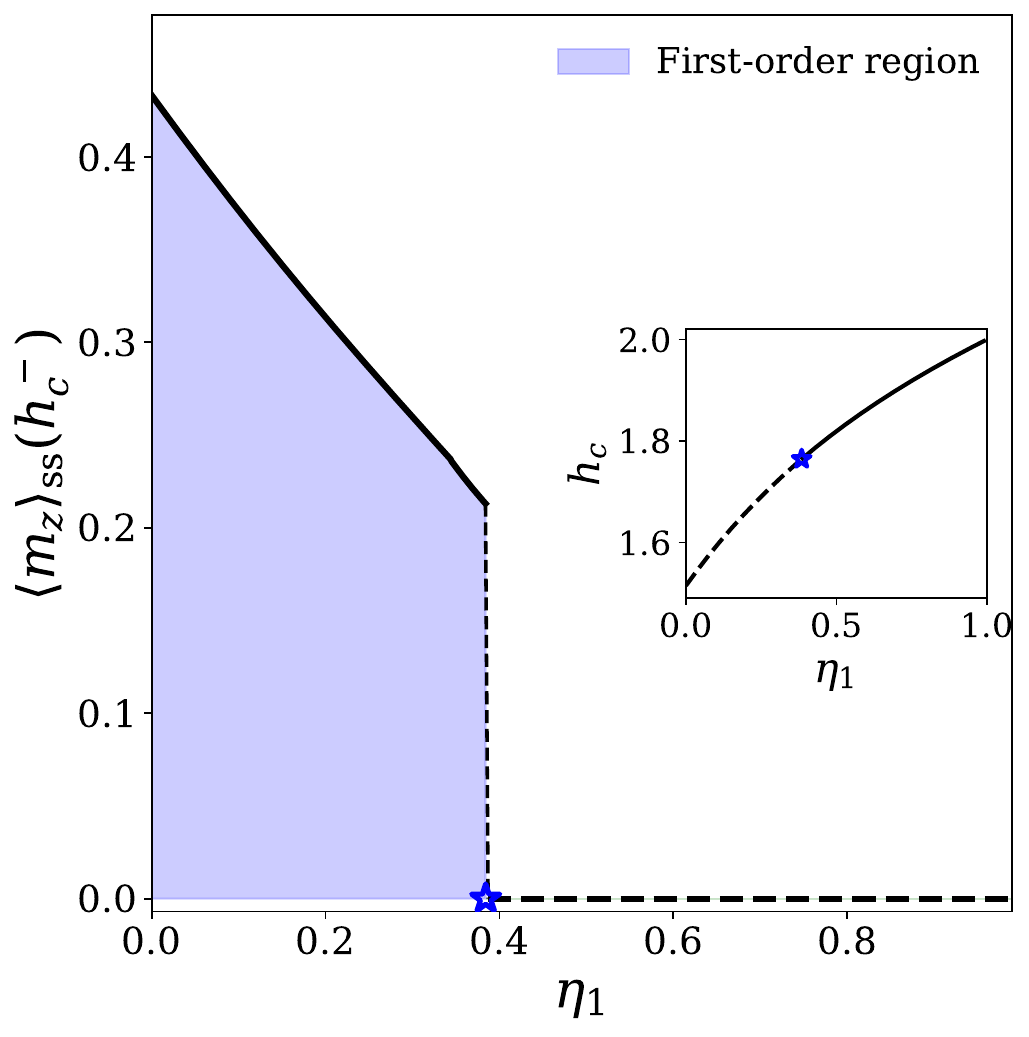}
\caption{Corresponding to the stationary state of the NLME~\eqref{eq:NLME_1}, here we show the quantity $\langle m_z \rangle_{\mathrm{ss}}(h_c^-)$, obtained by evaluating $\langle m_z \rangle_{\mathrm{ss}}$ just below the transition point $h_c$ by using Eq.~\eqref{eq:mz_eq_NLME}, as a function of the retention probability $\eta_1$, for fixed $\eta_2=0.5$, $K=0.22$, and $\beta J=2/3$. For the corresponding bare model ($\eta_1=\eta_2=0$), the transition is first order (see Fig.~\ref{fig:Fig1}). The inset shows the behavior of $h_c$ as a function of $\eta_1$ for the same choice of the parameters as in the main plot, where the nature of the transition changes from first order (black dashed) to continuous (black solid) across the tricritical point marked as the blue star. }
\label{fig:mz_jump_vs_eta1}
\end{figure}
The modification of the phase boundaries in the $(h,K)$-plane induced by the parameters $\eta_1, \eta_2$, as evidenced in Fig.~\ref{fig:NLME_kh_diagram}, may be unveiled more remarkably by investigating the behavior of $\langle m_z \rangle_{\mathrm{ss}}$ versus $h$  on changing $\eta_1$ while keeping fixed the set of parameters $J$, $K$, $\beta$, $\eta_2$. To properly quantify the nature of the phase transition as $h$ is varied, Fig.~\ref{fig:mz_jump_vs_eta1} shows as a function of $\eta_1$ the magnitude of $\langle m_z \rangle_{\mathrm{ss}}$ evaluated just below $h_c$,  denoted by the quantity $\langle m_z \rangle_{\mathrm{ss}}(h_c^-)$, along with the monotonic increase of transition point $h_c$ (see inset of Fig.~\ref{fig:mz_jump_vs_eta1}). For a fixed $\eta_2$, we observe that up to a specific value $\overline{\eta}_1$ of $\eta_1$, the system exhibits a first-order phase transition, which is characterized by a finite value of $\langle m_z \rangle_{\mathrm{ss}}(h_c^-)$. Beyond $\overline{\eta}_1$, however, the transition becomes continuous, and one has $\langle m_z \rangle_{\mathrm{ss}}(h_c^-)=0$. In the $(\eta_1,h_c)$-plane, the quantity $\overline{\eta}_1$ marks the location of a tricritical point, as shown in the inset of Fig.~\ref{fig:mz_jump_vs_eta1}. Thus, changing retention probability parameters results in not only shifting of the transition point, but also quite remarkably modifying the nature of the phase transition.

Altogether, the ability to tune the system's critical behavior via dissipative mechanisms highlights the rich phenomenology accessible within the studied non-equilibrium framework of imperfect quantum jumps between the system and the environment.

\section{Conclusion}\label{sec:conclu}
In this work, we presented a dynamic characterization of the generalized LMG model, demonstrating how the stationary-state phase diagram can be engineered and attained within an open quantum system framework. Within the standard Lindblad master-equation approach with thermally balanced quantum jump events, we first reproduced the model's canonical-equilibrium phase diagram. We then explored the consequences of a nonlinear Lindblad dynamics modeled in terms of imperfect jump events. By introducing two retention probability parameters $\eta_1$ and $\eta_2$ to model the corresponding imperfect jumps, we revealed a novel protocol for controlling the phase diagram, which is of nonequilibrium nature. The two jump-retention probabilities emerge as independent control parameters that reshape the phase diagram, not only shifting phase boundaries but also modifying the nature of the underlying phase transitions. This demonstrates that the interplay between coherent interactions and controlled dissipative processes can fundamentally alter the collective properties of long-range interacting quantum systems.

Beyond the generalized LMG model, our results illustrate how tailoring the linearity structure of Lindblad dynamics provides a systematic route to engineering stationary states that are inaccessible within conventional thermal equilibrium. We expect that the framework developed here can be extended to other interacting quantum many-body systems, and may serve as a useful guide for designing nonequilibrium phases through controlled dissipation in current quantum simulation platforms.

\section{Acknowledgements} AA acknowledges useful discussions with D. Arrufat Vicente. SG and AA acknowledge the generous allocation of computational resources of the Department of Theoretical Physics, TIFR, and the assistance of Kapil Ghadiali and Ajay Salve, and the financial support of the Department of Atomic Energy, Government of India under Project Identification No. RTI-4012.

\bibliography{manuscript}

\appendix
\section{Derivation of Eq.~\eqref{eq:NLME_1}}\label{sec:app1}
To derive Eq.~\eqref{eq:NLME_1}, we proceed as follows. We start with redefining the jump operators as $L'^{(\alpha)}\equiv\sqrt{\gamma^{(\alpha)}}L^{(\alpha)}$ with $\gamma^{(\alpha)}>0$. Given the state $|\psi(t)\rangle$ at time $t$, in the ensuing infinitesimal time interval $dt$, the state $|\psi(t+dt)\rangle$ is updated according to the following rule:
\begin{widetext}
    \begin{align}
    |\psi(t+dt)\rangle =\begin{cases} 
    \frac{(1 - i H_{\mathrm{eff}} dt)|\psi(t)\rangle}
{\sqrt{\langle \psi(t)|{(1 + i H_{\mathrm{eff}}^\dagger dt)(1 - i H_{\mathrm{eff}} dt)}|{\psi(t)}\rangle}}& \text{with probability } (1-P dt), \\
   \frac{L'^{(\alpha)} |\psi(t)\rangle}{\sqrt{\langle \psi(t)|L'^{(\alpha)\dagger}L'^{(\alpha)}|\psi(t) \rangle}} & \text{with probability } p_\alpha(1-\eta_\alpha)  dt ,\\
   |\psi(t)\rangle & \text{with probability }  p_\alpha \eta_\alpha dt ,
   \end{cases} \label{eq:psi_t_conditioned}
\end{align}
\end{widetext} 
where we have defined $p_\alpha\equiv \langle \psi(t)| L'^{(\alpha)\dagger} L'^{(\alpha)}|\psi(t) \rangle=\mathrm{Tr}[L'^{(\alpha)\dagger} L'^{(\alpha)}\rho]$, and $P\equiv\sum_\alpha p_\alpha$, while $H_\mathrm{eff}$ is given by $H_\mathrm{eff}= H - (i/2) \sum_\alpha L'^{(\alpha)\dagger} L'^{(\alpha)}$.

Noting that
$\langle \psi(t)|{(1 + i H_{\mathrm{eff}}^\dagger dt)(1 - i H_{\mathrm{eff}} dt)}|{\psi(t)}\rangle=1- P dt$, to leading order in $dt$, one may express the mixed state density matrix at time $t+dt$ as
\begin{align}
    &\rho(t+dt)\nonumber\\
    &=\frac{(1 - i H_{\mathrm{eff}} dt)\rho(t) (1 + i H^\dagger_{\mathrm{eff}} dt)}{1-P dt} (1-P dt)\nonumber\\
    &+\sum_\alpha \frac{L'^{(\alpha)}\rho(t) L'^{(\alpha)\dagger}}{p_\alpha}p_\alpha (1-\eta_\alpha)dt+\sum_\alpha \rho(t) p_\alpha \eta_\alpha dt\nonumber\\
    &=\rho(t)- i dt \left(H_{\mathrm{eff}} \rho(t)- \rho(t) H^\dagger_{\mathrm{eff}}\right)\nonumber\\
    &+dt \sum_\alpha \left( (1-\eta_\alpha) L'^{(\alpha)}\rho(t) L'^{(\alpha)\dagger} +\rho(t)p_\alpha \eta_\alpha\right),\label{eq:rho_t+dt_eq1}
\end{align}
where the last line is obtained by keeping terms up to $\mathcal{O}(dt)$. The expression of $H_\mathrm{eff}$ gives $-idt \left(H_{\mathrm{eff}} \rho(t)- \rho(t) H^\dagger_{\mathrm{eff}}\right)=-i [H, \rho]-\sum_\alpha \frac{1}{2}\{L'^{(\alpha)\dagger} L'^{(\alpha)}, \rho\}$.

Taking the limit $dt \to 0$ and putting $p_\alpha=\langle L'^{(\alpha)\dagger} L'^{(\alpha)} \rangle$, Eq.~\eqref{eq:rho_t+dt_eq1} may be written as
\begin{align}
\frac{d \rho}{dt}
&=-i [H, \rho]
+ \sum_\alpha \gamma^{(\alpha)}
\left(-\frac{1}{2} \{L^{(\alpha)\dagger} L^{(\alpha)}, \rho\} \right.\nonumber\\
&\quad\left.+ (1-\eta_\alpha) L^{(\alpha)} \rho L^{(\alpha)\dagger}+ \eta_\alpha \langle L^{(\alpha)\dagger} L^{(\alpha)} \rangle \rho\right). \label{eq:derived_NLME_full_rho}
\end{align}
The nonlinear evolution in Eq.~\eqref{eq:derived_NLME_full_rho} for the full density matrix $\rho$ maps directly onto the dynamics of the single-site density matrix $\rho_i$ under a mean-field structure of $\rho$ and the site-dependent additive structure of the jump channels, i.e., with $L^{(\alpha)} \to \sum_{i=1}^N L^{(\alpha)}_i$. Following the same steps that led from Eq.~\eqref{eq:Lindblad} to Eq.~\eqref{eq:MF-Lindblad}, one obtains using Eq.~\eqref{eq:derived_NLME_full_rho} the evolution of the single-site density matrix as
\begin{align}
\frac{d \rho_i}{d t}&= -i [H_{\mathrm{MF}}, \rho_i]+\sum_{\alpha}\gamma^{(\alpha)}\left(-\frac{1}{2}\{L^{(\alpha)\dagger}_i L^{(\alpha)}_i, \rho_i \}\right.\nonumber\\
&\qquad\left.+ (1-\eta_\alpha) L^{(\alpha)}_i \rho_i L^{(\alpha)\dagger}_i+\eta_\alpha \langle L^{(\alpha)\dagger}_i L^{(\alpha)}_i \rangle \rho_i\right), \label{eq:NLME_1_appendix}
\end{align}
where $\langle L_i^{(\alpha)\dagger} L_i^{(\alpha)} \rangle=\mathrm{Tr}[L_i^{(\alpha)\dagger} L_i^{(\alpha)}\rho_i ]$. The above is Eq.~\eqref{eq:NLME_1} of the main text.

\section{Locating first-order and continuous transition points from a self-consistency equation for the order parameter}\label{sec:app2}

Suppose we are given a transcendental self-consistency equation
\begin{align}
m=g(m)
\label{eq:m_gm}
\end{align}
for an order parameter $m$ that undergoes a phase transition as the dynamical parameters of the system (temperature, coupling, external field, etc.), on which $g$ depends, are varied. In the main text, the self-consistency equation considered, Eq.~\eqref{eq:mz_eq_NLME}, has $m$ as the magnetization, whose allowed range is $-1\le m\le1$. If the system possesses the symmetry $m\rightarrow -m$, as is the case for our model, the function $g(m)$ is odd and admits the expansion
\begin{align}
g(m)=\mathcal{A}m+\mathcal{B}m^3+\mathcal{C}m^5+\cdots,
\label{eq:m_series_expansion}
\end{align}
where the coefficients $\mathcal{A}$, $\mathcal{B}$, $\mathcal{C},\ldots$ depend on the dynamical parameters of the model. Owing to the $m\rightarrow -m$ symmetry, it is sufficient to restrict the discussion to the interval $0\le m\le1$. The solution $m^*$ of Eq.~\eqref{eq:m_gm} is determined by treating the equation as an iterative map, $m^{(n+1)}=g(m^{(n)})$, where $n$ denotes the level of iteration. The solution $m^*$ is then obtained as the result to which the iteration converges for large $n$, which requires the stability criterion $|g'(m^*)|< 1$~\cite{alligood1997chaos}.

\begin{figure}[h]
	\centering
	\includegraphics[width=\linewidth, height=0.5\linewidth]{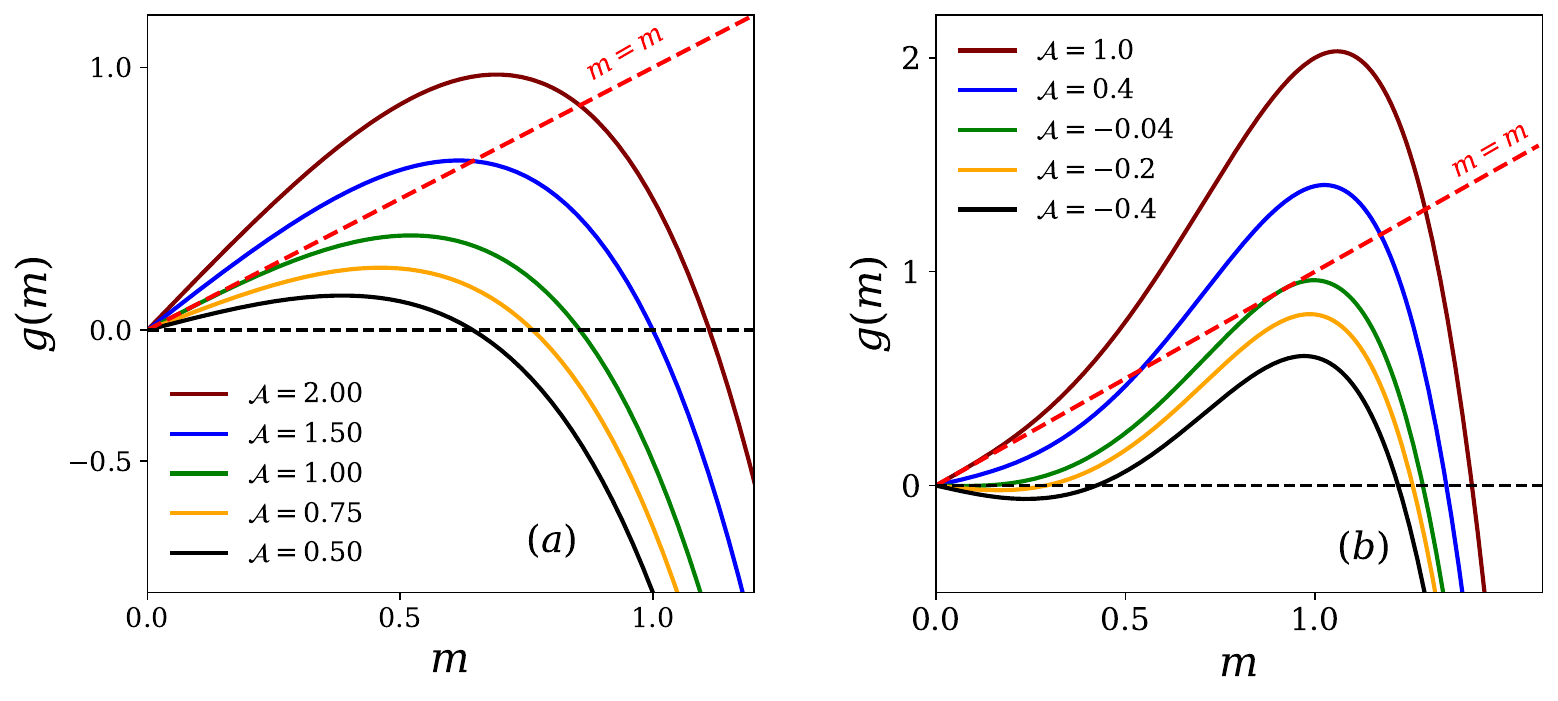}
	\caption{Behavior of the function $g(m)$ in Eq.~\eqref{eq:m_gm} for (a) the continuous transition regime ($\mathcal{B}<0$), and (b) the first-order transition regime ($\mathcal{B}>0$). In (a), one has (i) a zero stable solution for $\mathcal{A}<1$, and (ii) a non-zero stable solution for $\mathcal{A}>1$. In (b), one has (i) a zero stable solution for $\mathcal{A}<\mathcal{A}^*$, (ii) coexistence of two stable solutions, one zero and one non-zero, at $\mathcal{A}=\mathcal{A}^*$ (for the plot, we have $\mathcal{A}^*\approx -0.04$), and (iii) a non-zero stable solution for $\mathcal{A}>\mathcal{A}^*$. }
	\label{fig:G_function_behavior}
\end{figure}

Consider first the case $\mathcal{B}<0$. Here, for $\mathcal{A}<1$, one has only the zero solution $m^*=0$ of Eq.~\eqref{eq:m_gm}, which is stable since $|g'(0)| = \mathcal{A} < 1$. On the other hand, for $\mathcal{A}>1$, one has in addition a non-zero solution $m^*>0$, where one has $|g'(m^*)| < 1$, see Fig.~\ref{fig:G_function_behavior}(a). Then, the non-zero solution becomes stable and favored with respect to the zero solution. Consequently, one has the following scenario:  As $\mathcal{A}$ approaches unity from above, the non-zero solution continuously approaches zero and merges with the zero solution at $\mathcal{A}=1$. Then, the conditions $\mathcal{A}=1$, $\mathcal{B}<0$ identify a continuous phase transition.

Now, consider the case $\mathcal{B}>0$, and assume $\mathcal{C}<0$ so that at least one stable fixed point exists in the interval $0 \le m \le 1$. In this case, for $\mathcal{A}$ significantly below unity, there is only the zero solution $m^*=0$ of Eq.~\eqref{eq:m_gm}. As $\mathcal{A}$ increases, one obtains for a specific value $\mathcal{A}=\mathcal{A}^*$ the zero solution coexisting with a non-zero solution with both being stable: $|g'(0)|=|g'(m^*)|<1$. Beyond this point, two non-zero solutions appear, among which only one is stabilised by the criterion $|g'(m^*)|<1$. In such a situation, the conditions $\mathcal{A}=\mathcal{A}^*$, $\mathcal{B}>0$ locate the first-order transition line in parameter space, provided $\mathcal{C}<0$, see Fig.~\ref{fig:G_function_behavior}(b). The continuous and first-order transition lines meet at the tricritical point, characterized by $\mathcal{A}=1$, $\mathcal{B}=0$.

\end{document}